\documentclass[twocolumn, astrosymb]{aastex7}

\usepackage{lipsum}
\usepackage{breqn}
\usepackage{amsmath}
\usepackage{gensymb}
\usepackage{hyperref}
\usepackage{ulem}
\usepackage{longtable}

\newcommand{\HI}{\ion{H}{1}~}
\newcommand{\HII}{\ion{H}{2}~}

\newcommand{\Ha}{H$\alpha$~}
\newcommand{\ngc}{NGC 3344~}
\shorttitle{Linking Radial Gas Motion to ALM Regions}
\shortauthors{Olvera et al.}
\received{April 22, 2026}
\revised{August 21, 2026}
\accepted{September 15, 2026}
\submitjournal{ApJ}
\graphicspath{{./}{ngc3344_figs/}}

\begin{document}

\title{DIISC-VII: Linking Radial Gas Motions to Anomalously Low Metallicity Star-forming Regions in the XUV Disk of NGC 3344}

\correspondingauthor{Alejandro J. Olvera}
\email{ajolver2@asu.edu}

\author[0000-0002-2819-0753, sname=Olvera, gname=Alejandro]{Alejandro J. Olvera}
\affiliation{School of Earth and Space Exploration, Arizona State University, 781 Terrace Mall, Tempe, AZ 85287, USA}
\email{ajolver2@asu.edu} 

\author[0000-0002-2724-8298, sname=Borthakur, gname=Sanchayeeta]{Sanchayeeta Borthakur}
\affiliation{School of Earth and Space Exploration, Arizona State University, 781 Terrace Mall, Tempe, AZ 85287, USA}
\email{sborthak@asu.edu}

\author[0000-0003-4019-0673, sname=Di Teodoro, gname=Enrico]{Enrico M. Di Teodoro}
\affiliation{Università di Firenze, Dipartimento di Fisica e Astronomia, via G. Sansone 1, 50019 Sesto Fiorentino, Firenze, Italy}
\affiliation{INAF - Osservatorio Astrofisico di Arcetri, Largo E. Fermi 5, I-50125 Firenze, Italy}
\email{enrico.diteodoro@unifi.it}

\author[0000-0002-8768-9298, sname=Saha, gname=Kanak]{Kanak Saha}
\affiliation{Inter-University Centre for Astronomy and Astrophysics, Ganeshkhind, Post Bag 4, Pune 411007, India}
\email{kanak@iucaa.in}

\author[0000-0002-3472-0490, sname=Padave, gname=Mansi]{Mansi Padave}
\affiliation{Department of Astronomy $\&$ Astrophysics, University of California, San Diego, 9500 Gilman Drive, LaJolla, CA 92093, USA}
\email{mpadave@ucsd.edu}

\author[0000-0003-1436-7658, sname=Gim, gname=Hansung]{Hansung B. Gim}
\affiliation{Eureka Scientific Inc., 2452 Delmer Street Suite 100, Oakland, CA 94602, USA}
\email{hansung.b.gim@gmail.com}

\author[0000-0003-3168-5922, sname=Momjian, gname=Emmanuel]{Emmanuel Momjian}
\affiliation{National Radio Astronomy Observatory, P.O. Box O, Socorro, NM, USA}
\email{emomjian@nrao.edu}

\begin{abstract}
We present evidence connecting anomalously low-metallicity regions to radial gas inflows in the extended ultraviolet (XUV) disk of the galaxy NGC~3344. Anomalously low-metallicity \ion{H}{2} regions are hypothesized to result from the accretion of metal-poor gas, but a direct link between gas flows and low metallicity gas has not been previously established. We use high-resolution \ion{H}{1}-21cm imaging from the Very Large Array to trace neutral gas kinematics and multi-slit optical spectroscopy from the MMT to trace the gas-phase metallicity of 76 \ion{H}{2} regions in NGC~3344. The galaxy exhibits a negative radial metallicity gradient of -0.378~dex~$\rm R_{25}^{-1}$, with \ion{H}{2} regions in the XUV disk showing nearly twice the metallicity scatter of those in the inner disk and several showing an anomalously low metallicity. Modeling the \ion{H}{1} disk as tilted rings with $\rm ^{3D}BAROLO$, we find gas moving radially inwards within the XUV disk at an average radial velocity of 6 $\rm km~s^{-1}$, likely fueling star formation and diluting the metallicity, thereby producing anomalously low metallicity regions. Chemical evolution models indicate that the gas fraction, effective yield, and mass-loading factor profiles are disrupted at the onset of the XUV disk, as expected from recent gas flows. This example is the first direct detection of how radial gas flows can support stellar disk expansion and cause inside-out galaxy growth.
\end{abstract}



\section{Introduction} \label{sec:intro}
Star-forming galaxies with extended ultraviolet (XUV) disks exhibit UV-bright emission well beyond their optical boundaries and make up about 30\% of late-type galaxies in the local Universe \citep{GildePaz_2005ApJ...627L..29G, Thilker_2007ApJS..173..538T}. These galaxies provide an excellent laboratory for studying star formation in outer disks. A prominent example is NGC 3344, whose XUV disk spans an area three times larger than its inner optical disk. \cite{Padave_2021ApJ...923..199P} showed that star formation occurs across the extended region. From the inner to the outer edge of the XUV disk, the specific star formation rate (sSFR) rises from $\rm 10^{-10}~yr^{-1}$ to $\rm 10^{-8}~yr^{-1}$, indicating starburst-like activity capable of doubling the stellar mass in the outskirts within about 0.5 Gyr. Such extended bursts can be triggered by mechanisms such as gas compression due to spiral density waves \citep{Bush_2010ApJ...713..780B} or by inflows of fresh gas refueling gas reservoirs \citep{Mihos_1996ApJ...464..641M, Dekel_2023MNRAS.523.3201D}. Galaxy interactions and minor mergers can introduce fresh gas into a galaxy's ecosystem; however, NGC 3344 shows no signs of recent interactions or mergers.

To sustain star formation for longer than a few Gyrs, galaxies must continuously accrete gas from the circumgalactic medium (CGM) to replenish their gas reservoirs \citep{Keres_2005MNRAS.363....2K,Bigiel_2008AJ....136.2846B,Hafen_2022MNRAS.514.5056H}.
Yet the mechanisms governing the migration of CGM gas into disks and how this fresh material influences the kinematics and metallicity of the existing interstellar medium (ISM) remain poorly constrained. Directly detecting inflows, whether onto or within the disk, is notoriously difficult due to the low gas densities and small radial velocities involved. Extended \ion{H}{1} disks, present in more than 70\% of star-forming galaxies, offer a potentially efficient channel for transporting gas inward from the outskirts toward the central regions \citep{Broeils_1997A&A...324..877B,Swaters_2002A&A...390..829S,Noordermeer_2005A&A...442..137N}. Observations from The \HI Nearby Galaxy Survey (THINGS; \citealt{Walter_2008AJ....136.2563W, Schmidt_2016MNRAS.457.2642S}) revealed radial velocities on the order of $\rm \sim10~km~s^{-1}$ or less . However, more recently, \cite{DiTeodoro_2021ApJ...923..220D} measured low radial velocities within the \HI disks of 54 nearby star-forming galaxies and concluded that there is no systematic evidence for radial inflows. Thus, the question of radial inflows within disks remains highly debated. 

Independently, chemical signatures can be used to investigate gas flows. Inflows can leave indirect signatures by introducing chemically distinct gas into the native ISM. When fresh, metal-poor gas is integrated into the disk, it can dilute \HII regions, either through vertical infall onto the disk or through gas moving laterally within it, causing these regions to deviate from the galaxy's underlying radial metallicity gradient. Diluted regions such as these were coined anomalously low metallicity (ALM) regions by \cite{Hwang_2019ApJ...872..144H}, who found that 25\% of the star-forming galaxies in the Mapping Nearby Galaxies at Apache Point Observatory (MaNGA; \citealt{Bundy_2015ApJ...798....7B}) Survey host ALM regions. \cite{Olvera_2024ApJ...976..205O} discovered ALM \HII regions within the mid-disk of NGC~99 and showed that the gas mass required to dilute these regions to their present-day metallicities is comparable to that of the Milky Way’s high-velocity clouds. A variety of observational and theoretical works support the idea that ALM regions arise from dilution by low-metallicity gas arriving into the disk \citep{Howk_2018ApJ...856..166H, Luo_2021ApJ...908..183L, Wang_2022ApJ...938L..16W, Ju_2022ApJ...938...96J, Grossi_2025arXiv250118498G, Long_2025ApJ...986...31L}. 

If gas from the CGM enters the extended \ion{H}{1} disk and migrates inwards, it could simultaneously fuel widespread star formation across the entire outer disk and dilute multiple star-forming regions, thereby creating ALM \HII regions. Numerous simulations \citep{Keres_2005MNRAS.363....2K,Dekel_2009Natur.457..451D, Keres_2009MNRAS.395..160K, Stewart_2011ApJ...738...39S, Sankar_2025arXiv251103793S} and analytical studies \citep{Fraternali_2012MNRAS.426.2166F, Saha_2014MNRAS.444..352S, Bland-Hawthorn_2017ApJ...849...51B} support this picture, suggesting that gas inflows into extended \ion{H}{1} reservoirs can help drive outer-disk star formation. Observational evidence likewise hints at a link between outer-disk star formation, extended \ion{H}{1} structures, and CGM-fed gas accretion \citep{Bigiel_2010AJ....140.1194B,Borthakur_2015ApJ...813...46B}. However, a more direct connection between outer-disk inflows and ongoing star formation and chemical evolution remains to be established.

In this paper, we present, for the first time, the missing link between low-metallicity regions and radial gas inflows in the XUV disk galaxy NGC~3344. In Section \ref{sec:obs and data}, we describe why  \ngc is best suited for this study, present new optical spectroscopy of its star-forming regions, and detail how we measure radial velocities. In Section \ref{sec:results}, we present our measurements. Section \ref{sec:discussion} discusses offsets we observe in the radial profiles of various properties of NGC 3344. We summarize our results and leave our final conclusions in Section \ref{sec:conclu}. Throughout the rest of paper, we use the following shorthand notation to symbolize integrated emission line fluxes: [\ion{N}{2}] $\equiv$ [\ion{N}{2}]$\lambda6584$; [\ion{S}{2}] $\equiv$ [\ion{S}{2}]$\lambda6717$+[\ion{S}{2}]$\lambda6731$; and [\ion{O}{3}] $\equiv$ [\ion{O}{3}]$\lambda5007$.

\section{Observations \& Methods} \label{sec:obs and data} 

\begin{figure}[h!]
\includegraphics[width =\linewidth]{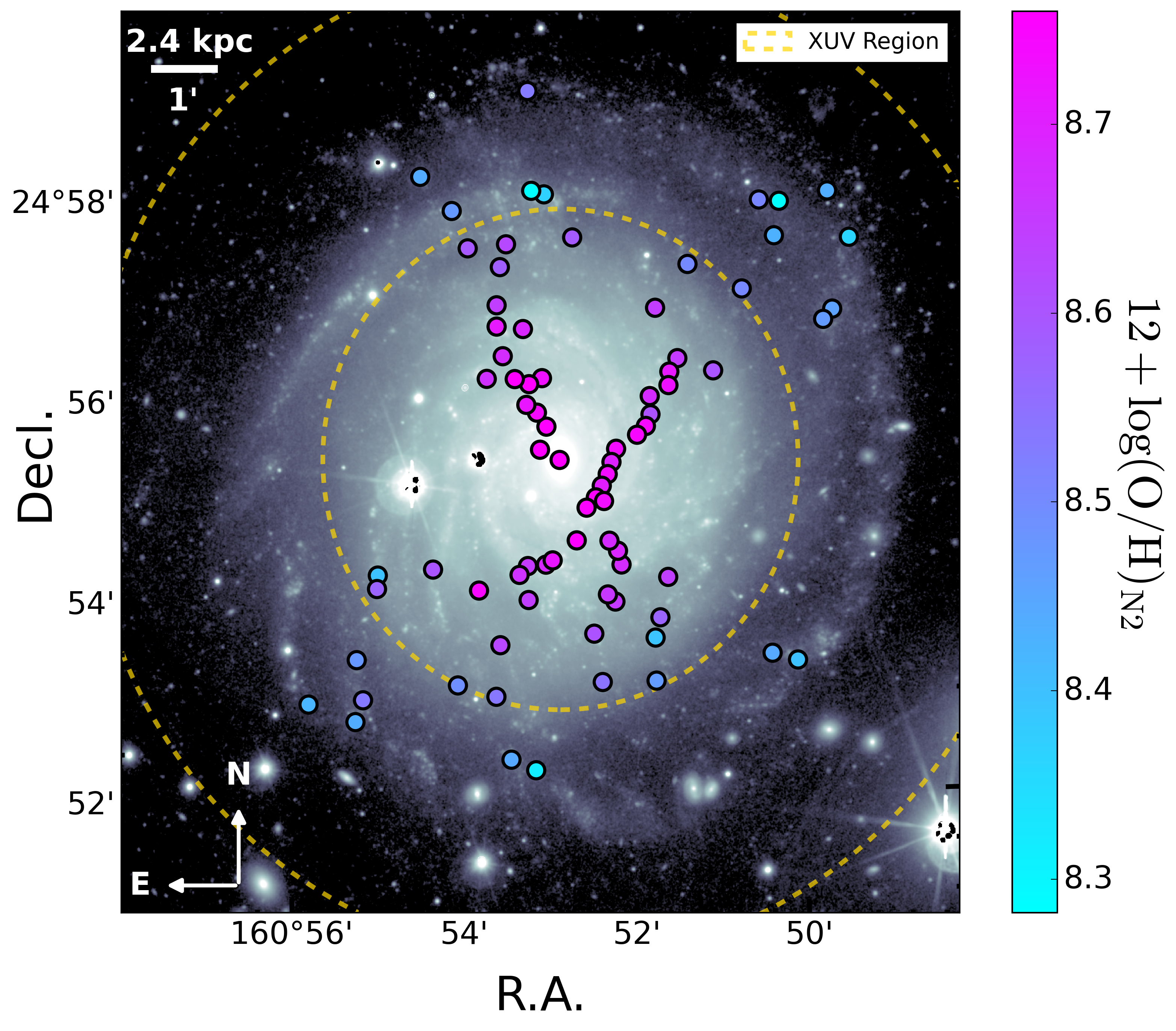}
\caption{A r-band image of NGC 3344 obtained with the Vatican Advanced Technology Telescope is shown. The circles, color-coded by their gas-phase metallicity, represent the locations of our MMT/Binospec slits. The yellow dashed lines indicate the XUV region, which starts at 6 kpc and ends at 12 kpc.}
\label{fig:HII_regions}
\end{figure}

\subsection{NGC 3344}\label{subsec:ngc3344}
\ngc is a local star-forming galaxy with a Type-I XUV disk starting at 6 kpc and shows no signs of tidal disruptions \citep{Thilker_2007ApJS..173..538T,Padave_2021ApJ...923..199P}. The galaxy has a star-formation rate (SFR) of $\rm log(SFR/M_\odot ~yr^{-1}) = -0.33$, a stellar mass ($\rm M_\star$) of $\rm log(M_\star/M_\odot ) = 10.18$, and a \HI mass ($\rm M_{HI}$) of $\rm log(M_{HI}/M_\odot ) = 9.44$ \citep{Padave_2021ApJ...923..199P,Padave_2025ApJ...986..145P}. Additionally, \ngc is a part of the Deciphering the Interplay between the Interstellar medium, Star, and Circumgalactic medium (DIISC) survey \citep{Borthakur_2024arXiv240912554B}, which combines ultraviolet spectroscopic data from the Cosmic Origins Spectrograph aboard the Hubble Space Telescope with \HI 21 cm hyperfine transition high-resolution imaging from the NSF's Karl G. Jansky Very Large Array (VLA), ultraviolet imaging from Galaxy Evolution Explorer (GALEX), and optical imaging from the Vatican Advanced Technology (VATT). DIISC data products include stellar and \HI surface density maps, SFR and dust maps, and absorption line spectroscopy. With this suite of data available, NGC 3344 is the perfect candidate to identify a direct connection between gas inflows and star formation within the outer disk. The UV-bright structures in the XUV disk indicate recent, ongoing star formation, thus making it simple to probe the metallicities of the \HII regions. The extended \HI disk of the galaxy is also easily detected in VLA imaging, allowing us to apply kinematic models to test for inflows. For this study, we adopt a distance of 8.28 kpc \citep{Sabbi_2018ApJS..235...23S}, which corresponds to a linear scale of 40 pc/$^{\prime\prime}$. The close proximity of the galaxy yields access to high-resolution spatial data: 1$^{\prime\prime}$ slits for the Binospec spectroscopy, an average PSF of 1$^{\prime\prime}$ for the VATT imaging, and a PSF of 3$^{\prime\prime}$ for the GALEX imaging, and an average beam size of 6$^{\prime\prime}$ for the VLA HI imaging.

\subsection{MMT Binospec Observation of HII Regions}
In addition to the DIISC data discussed in Section \ref{subsec:ngc3344}, we obtained new multi-object slit spectra of 85 H$\alpha$-bright regions in \ngc using the optical spectrograph, Binospec, on the 6.5m MMT telescope located on Mt. Hopkins in Arizona. The location of each slit is shown in Figure \ref{fig:HII_regions} and given in Table \ref{tab:properties}. Observations were split across two nights: 46 regions were observed on UT 2024 March 18, and the remaining 39 on UT 2024 May 8. Seeing ranged between $1.0^{\prime\prime}$ and $1.5^{\prime\prime}$ on both nights. Our spectroscopic targets were selected using \texttt{Source Extractor} \citep{Bertin_1996A&AS..117..393B} on an \Ha+[\ion{N}{2}] image of \ngc taken with the 1.8 m  Vatican Advanced Technology Telescope (VATT). The image and selection method are presented in \cite{Padave_2025ApJ...986..145P}. The \texttt{BinoMask} software \citep{kansky_2019PASP..131g5005K} was used to automatically select targets from the catalog by optimizing the number of targets per mask design. Each mask was observed for a total exposure time of 2400 seconds, divided into four 600-second exposures. We used Binospec’s 270 lines mm$^{-1}$ grating centered at 5800 \AA, providing a dispersion of 1.3 Å pix$^{-1}$ and a resolution power of R=1340. This setup enabled complete coverage of all relevant emission lines within the wavelength range of 3500–9240 \AA. The data were reduced using the Binospec Data Reduction Pipeline, which performed bias correction, flat-fielding, wavelength calibration, relative flux correction, co-addition, and extraction to 1D spectra \citep{kansky_2019PASP..131g5005K}.

The identification of emission lines and their flux measurement were carried out using the procedure described in \cite{Olvera_2024ApJ...976..205O}, with one important modification. Prior to measuring emission-line fluxes, the underlying stellar continuum of each extracted spectrum was modeled and subtracted using the Penalized Pixel-Fitting (\texttt{pPXF}) method \citep{Cappellari_2023MNRAS.526.3273C}. We used the E-MILES simple stellar population templates \citep{Vazdekis_2016MNRAS.463.3409V} distributed with \texttt{pPXF}, which are based on Padova isochrones and assume a \cite{Salpeter_1955ApJ...121..161S} initial mass function. Subtracting the best-fitting stellar continuum minimizes the effects of underlying stellar Balmer absorption, enabling more accurate measurements of the nebular emission-line fluxes. Emission-line fluxes were subsequently measured by fitting Gaussian profiles to the continuum-subtracted spectra, and only lines detected with a signal-to-noise ratio greater than three were retained for subsequent analysis. The observed spectroscopic H$\alpha$ fluxes and the Balmer decrements of the individual regions, are presented in Table~\ref{tab:properties}.

\subsection{Tilted Ring Fitting of VLA Data with 3D-Barolo}\label{subsec:HIdata}
This work utilizes high-resolution \HI 21 cm VLA observations in the D-, C-, and B-configurations obtained as part of the DIISC program and presented by \cite{Padave_2021ApJ...923..199P}. The synthesized beam of 6\farcs7$\times$5\farcs4 yields high spatial resolution comparable to that of the optical and UV data, as well as high spectral resolution (R = 29106). We use the 3D kinematical software $\rm ^{3D}BAROLO$\footnote{Code is available at \url{https://editeodoro.github.io/Bbarolo}. This paper uses
version 1.7.} \citep{DiTeodoro_2015MNRAS.451.3021D} to model the disk as a set of concentric tilted rings, allowing us to determine circular and radial velocities within the XUV disk of NGC 3344. Each ring is described by four geometrical parameters: the center of the galaxy ($x_0$, $y_0$), the position angle of the major axis of the disk ($\phi$), and the inclination of the disk with respect to the line of sight (e.g., $i=90\degree$ for edge-on). The rings also have four kinematic parameters: the systematic, rotational, and radial velocities ($\rm V_{sys}, V_{rot}, V_{rad}$) and the gas intrinsic velocity dispersion ($\sigma_{{\rm gas}}$). 

Our modeling follows the multi-step approach outlined in \cite{DiTeodoro_2021ApJ...923..220D}. We adopt the galaxy center from optical imaging, which provides higher spatial resolution than the \ion{H}{1} data, and set the ring spacing to 6$^{\prime\prime}$ to match the beam size of the \HI data. In the initial fit, $x_0$, $y_0$, $\rm V_{sys}$, and $\rm V_{rad}$ are fixed while the remaining parameters are allowed to vary. We use the same initial conditions from \cite{DiTeodoro_2021ApJ...923..220D}. Masks for the fit are generated using the built-in $\rm SEARCH$ source-finding algorithm, with primary and secondary thresholds set to 5$\sigma_{\rm rms}$ and 3$\sigma_{\rm rms}$, respectively, where $\sigma_{\rm rms}$ is the rms noise of the data.

As in \cite{DiTeodoro_2021ApJ...923..220D}, the inclination and position angle are regularized to a Bezier curve and only $\rm V_{rot}$ and $\sigma_{{\rm gas}}$ are allowed to vary. Additionally, we manually adjust the inclination of each ring after 11.2 kpc to decrease linearly with radius in order to keep the rotation curve flat in the outer disk. For the first two steps, a $(\cos\theta)^{2}$ weighting is applied to the fit, where $\theta$ is the azimuthal angle into the plane of the disk, since the information on $\rm V_{rot}$ lies around the major axis (e.g., $\theta\sim0\degree$). To measure the radial velocities, we fix all parameters to the values obtained in the first two fits, switch the weighting to $(\sin\theta)^{2}$, and allow only $\rm V_{rad}$ to vary. Interpreting the sign of $\rm V_{rad}$ as radial motions towards or away from the center of the galaxy requires knowledge of the rotation direction of the galaxy. \ngc rotates clockwise, meaning that positive (negative) $\rm V_{rad}$ values correspond to radial motions towards (away from) the center of the galaxy following $\rm ^{3D}BAROLO$'s definition.

\subsection{Calculating Stellar and Gas Masses}
We utilize a stellar mass surface density map created by \cite{Padave_2024ApJ...960...24P} using VATT optical continuum g- and r- band imaging and a CO moment-0 image created by \cite{Helfer_2003ApJS..145..259H} using millimeter observations from the Berkeley-Illinois-Maryland Association (BIMA) radio telescope array. CO intensities were converted to molecular gas masses by cross-matching fluxes with the total molecular gas mass from the Instituto de Radioastronomía Milimétrica (IRAM) telescope reported by \citet{Lisenfeld_2011A&A...534A.102L}. We define the gas mass to be $\rm M_{gas} \equiv 1.345(M_{HI}+M_{H_2}) = 1.345M_H$, where the multiplicative factor accounts for He. To avoid spatial decorrelation between cold molecular gas and \HII regions \citep{Schinnerer_2019ApJ...887...49S,Semenov_2021ApJ...918...13S}, we create $0.8\times0.8$ kpc masks centered at the position of each \HII region and integrate the pixels contained within each mask using \texttt{photutils} \citep{Bradley_larry_bradley_2023_7946442}. The masks are used to obtain the total $\rm M_{\star}$, \HI intensity, and $\rm M_{H_2}$ for each region. \HI intensity is then converted to \HI mass, or M$_{\rm HI}$, by multiplying the \HI surface density by the area of each region. 

\section{Results}\label{sec:results}

\subsection{Radial Metallicity Gradient}\label{sec:radial Z grad}
We examine how the gas-phase metallicity varies as a function of galactocentric radius in NGC~3344 using the strong-line method. We measure the integrated flux of ${\rm{H}}\alpha$, ${\rm{H}}\beta$, and [\ion{N}{2}] to estimate metallicity with the N2 indicator, which has the benefit of being unaffected by dust attenuation due to the proximity of the emission lines in wavelength space \citep{Searle_1971ApJ...168..327S,Alloin_1979A&A....78..200A,Kewley_2002ApJS..142...35K,Marino_2013A&A...559A.114M,Curti_FMR_2020MNRAS.491..944C}. The indicator is defined to be N2 = log([\ion{N}{2}]/H$\alpha$), where the line names symbolize the integrated flux of the line. N2 is converted to oxygen gas-phase metallicity $12 + \rm log(O/H)$ using the calibration from \cite{Curti_FMR_2020MNRAS.491..944C}. Each region's N2 and $12 + \rm log(O/H)$ value is given in Table \ref{tab:properties}. We note that our results do not change when utilizing the O3N2 metallicity indicator, which is defined to be log(\{[\ion{O}{3}] H$\alpha$\}/\{H$\beta$ [\ion{N}{2}]\}).

\begin{figure*}[]
\centering
    \includegraphics[width = 0.75\linewidth]{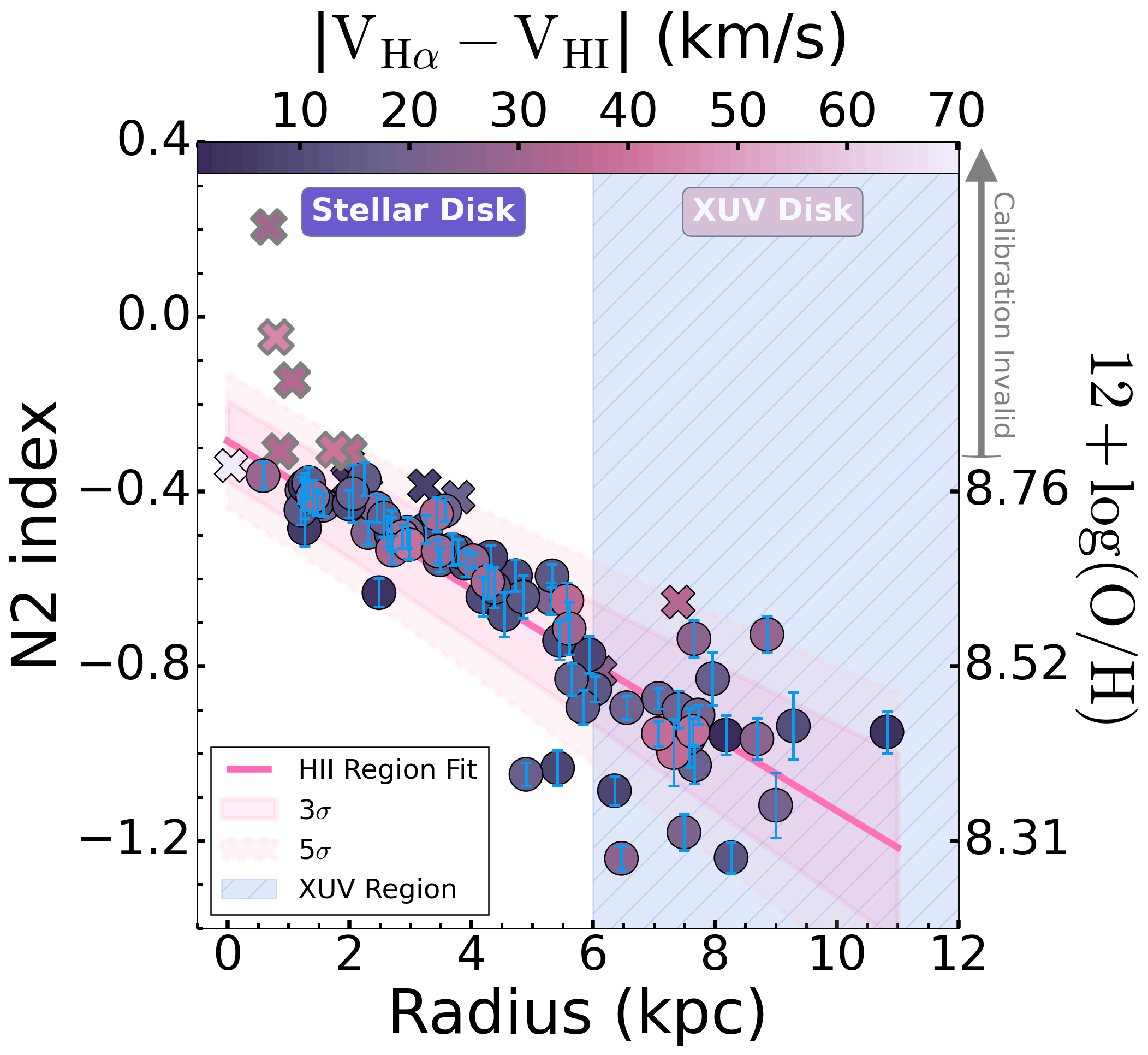}
\caption{The radial metallicity of NGC 3344 from the N2 indicator is shown as a function of galactocentric radius. The right-side y-axis shows the corresponding gas-phase metallicity. The calibration used to convert from the N2 indicator to $12 + \rm log(O/H)$ is invalid when N2 $>$ $-0.33$, which we indicate with the gray vertical arrow. The solid pink line represents the linear fit to the \HII region data (circles), while the dark pink and light pink shaded regions indicate the $3\sigma$ and $5\sigma$ confidence intervals of the fit, respectively. The colorbar shows the absolute difference between the ${\rm{H}}\alpha$ emission line velocity offset and the mass-weighted HI 21\,cm velocity of the ISM in the region. The blue-shaded hatch region indicates where the XUV region begins. Cross-shaped data points are regions that are not ionized by the UV radiation from young and massive stars.}
\label{fig:metallicity_gradient}
\end{figure*}

The radial metallicity gradient of NGC 3344 from the N2 indicator is shown in the top panel of Figure \ref{fig:metallicity_gradient} with the corresponding gas-phase metallicity given on the right side. We note that the calibration from \cite{Curti_FMR_2020MNRAS.491..944C} becomes invalid when N2 $>$ $-0.33$, so regions without a corresponding gas-phase metallicity value are indicated by a gray outline. Additionally, non-star-forming regions (based on the BPT diagram in Appendix \ref{sec:BPT}) are symbolized by the X symbol and excluded from any further analysis. In Figure \ref{fig:metallicity_gradient}, the solid pink line shows the best linear fit between N2 and galactocentric radius, with the $3\sigma$ and $5\sigma$ confidence intervals represented by the dark pink and light pink shaded regions, respectively. The data points are color-coded by the absolute difference between the ${\rm{H}}\alpha$ emission line velocity offset and the mass-weighted \HI 21\,cm velocity of the ISM in the region. The hatched blue region indicates the XUV region.

We use the package \texttt{lmfit} \citep{Newville_2021zndo....598352N} to find the best linear fit between $12 + \rm log(O/H)$ and galactocentric radius. NGC 3344 has a negative metallicity gradient of $-0.047\pm0.003$ dex kpc$^{-1}$ or $-0.378\pm0.023$ dex/$\rm R_{25}$. The fit yields a Spearman-$\rho$ of -0.92, which indicates a strong inverse correlation between metallicity and galactocentric radius. Regions within 6 kpc exhibit a significantly smaller offset from the fit than regions within the XUV disk of NGC 3344. The metallicity in the XUV disk widely varies, ranging from 8.28 to 8.55 dex. At the onset of the XUX disk, we note that several regions show more than 3$\sigma$ deviation from the mean and are identified as ALM regions.

\subsection{Detecting Radial Flows}\label{sec:tilted_ring}

\begin{figure*}[t]
\centering
    \includegraphics[width = \linewidth]{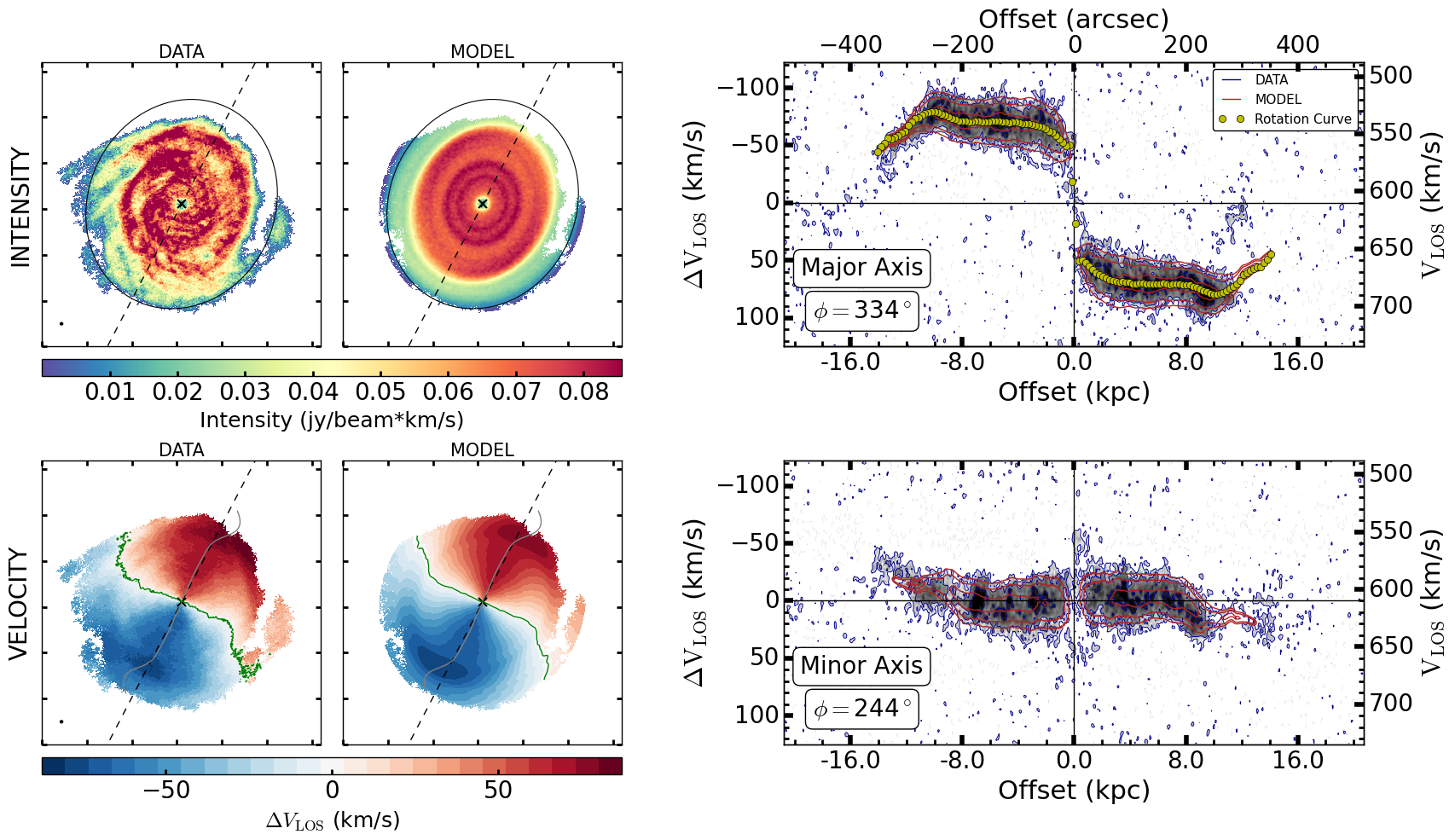}
\caption{The modeling of NGC 3344 with $\rm ^{3D}BAROLO$ is shown. \textit{Left}: The HI 21 cm intensity and line-of-sight velocity data and their best-fit model is portrayed. The 6\farcs7 $\times$ 5\farcs4 synthesized beam is shown on the lower left. The solid black ellipse has a semi-major and semi-minor axes length of 351$^{\prime\prime}$ and 310$^{\prime\prime}$, respectively. The gray line indicates the major axis of the galaxy, while the green line shows the systemic velocity. \textit{Right}: P-V slice through the average major (top) and minor (bottom) axes. The blue and red contour levels represent the data and model, respectively, at  $2^{n}\times\sigma_{rms}$ where $n=0\ldots5$ with $\sigma_{rms}=0.5$ mJy beam$^{-1}$. At $\sim$10 kpc, the line-of-sight velocity deviates from the systemic velocity in the minor axis plot, indicating a warp in the outer disk and the presence of radial flows.}
\label{fig:BBarolo_model}
\end{figure*}

We present the integrated \HI intensity (moment 0) and intensity-weighted velocity (moment 1) maps and best-fit models of NGC 3344 in the left column of Figure \ref{fig:BBarolo_model}. The right column of the figure shows position-velocity (P-V) slices through the \HI data, with the top and bottom panels showing slices averaged along the major and minor axes, respectively. The model, shown by the red contours, closely follows the \HI emission in the major-axis slice. At $\sim$8 kpc, the emission starts to deviate from the relatively flat rotation curve seen preceding this point. In the minor axis slice, the outer ends of the emission at 8 kpc are ``twisting" away from the systematic velocity. The cause of this deviation in both the slices is due to change in the inclination and position angle beyond 8 kpc, which was measured via the tilted ring modeling. Changes in inclination and position angle are an indication of a warped \HI disk.  As noted in \cite{DiTeodoro_2021ApJ...923..220D}, a position angle and inclination warp and radial gas motions can cause these twisting features to occur in the P-V slices. We discuss the implications of a warped outer \HI disk in Section \ref{subsec:linking radial flows and metallicity}.

We use $\rm ^{3D}BAROLO$ to quantify the radial velocities of the outer disk gas. The cyan points in the top panel of Figure \ref{fig:Vrad_deltaOH_A1} show the averaged radial velocity of each tilted ring from $\rm ^{3D}BAROLO$ as a function of galactocentric radius. The inner region of the galaxy ($< 3$ kpc) shows little radial gas motion. While there is a small amount of radial motion from 3 kpc to 4 kpc, the radial velocities are on average about $\rm -6~km~s^{-1}$ starting at 6 kpc, where the XUV disk also begins, indicating that gas is moving radially inwards towards the galactic center parallel to the disk. The magnitude of the radial velocities in the XUV disk agrees with values from previous larger studies such as \cite{Schmidt_2016MNRAS.457.2642S} and \cite{DiTeodoro_2021ApJ...923..220D}. At around 9.5 kpc, the radial velocities rapidly turn positive. Representative uncertainties of approximately 5 $\rm km~s^{-1}$ apply to the radial velocity measurements. See Appendix \ref{vrad_err} for a discussion of the uncertainty estimation in $\rm ^{3D}BAROLO$ and its limitations.

Following the prescription from \cite{DiTeodoro_2021ApJ...923..220D}, we calculate an average neutral gas flow rate as a function of radius, $\rm \dot{M}$, using 
\begin{equation}\label{equ:mass_flow}
    {\rm \dot{M}~(R)} = 1.33\cdot2 \pi {\rm R\Sigma_{HI}(R)V_{rad}(R)},
\end{equation}

\noindent where $\rm \Sigma_{HI}$ is the face-on \HI mass surface-density. The gas flow rate, shown by the orange points in Figure \ref{fig:Vrad_deltaOH_A1}, stays mostly neutral until the start of the XUV disk, where the rate reaches negative values of 4 $\rm M_\odot~yr^{-1}$, indicating gas inflow (or accretion) out to about 10 kpc.

\section{Discussion} \label{sec:discussion}

\subsection{Triggers of XUV Disks: Evidence from Radial Flows \& Metallicity Deviations}\label{subsec:linking radial flows and metallicity}
We observe both a warp in the outer H I disk (Section \ref{sec:tilted_ring}) and coherent inward radial motions beginning at the onset of the XUV disk. Warped disks are commonly interpreted as signatures of gas accretion with a misaligned angular momentum vector relative to the pre-existing disk, preventing the gas from immediately settling into equilibrium \citep{Binney_1992ARA&A..30...51B,Ostriker_1989MNRAS.237..785O,Jiang_1999MNRAS.303L...7J,Lopez_2002A&A...386..169L}. The resulting misalignment introduces torques and dissipative interactions between gas layers, leading to angular momentum loss and driving radial inflow \citep{Pringle_1992MNRAS.258..811P, Ogilvie_1999MNRAS.304..557O, Sancisi_2008A&ARv..15..189S, DiTeodoro_2021ApJ...923..220D}. In this scenario, gas accreted into the outer H I disk is transported inward, naturally linking the observed warp to the non-circular motions seen in the P–V slices. 

\begin{figure}[t]
\centering
    \includegraphics[width = \linewidth]{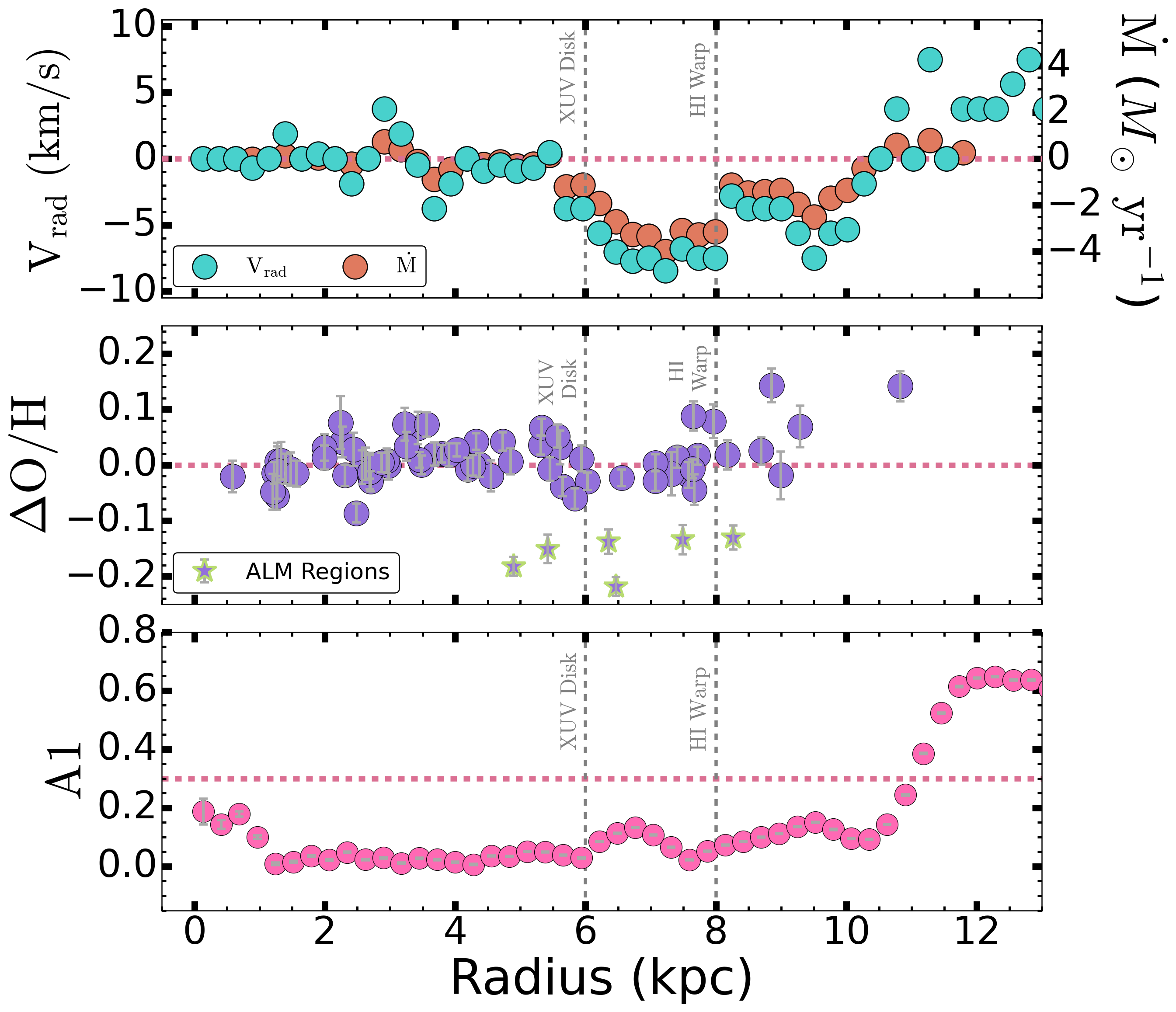}
\caption{(\textit{top}) The radial profile of each ring's radial velocity and mass flow rate is shown in cyan and orange, respectively, with positive/negative values indicating gas moving radially outwards/inwards. Representative uncertainties of approximately 5 $\rm km~s^{-1}$ apply to the radial velocity measurements (see Appendix \ref{vrad_err} for further discussion). (\textit{middle}) Each region's metallicity deviation as a function of galactocentric radius. Grey error bars represent the propagated uncertainties on the metallicity measurements. At the onset of the XUV disk, radial velocities become negative, and the measured metallicities deviate markedly from expectations. ALM regions are symbolized by stars. (\textit{bottom}) The pink points indicate the m=1 Fourier amplitude of each ring. Error bars represent the uncertainties in the measured Fourier amplitudes. At A1$>0.3$, a disk is considered strongly lopsided.}
\label{fig:Vrad_deltaOH_A1}
\end{figure}

If this inward-moving gas originates in the outer disk or the CGM, it is expected to be lower in metallicity than the pre-existing ISM. As this gas mixes with the ISM, it dilutes the metallicity of star-forming regions, causing them to deviate from the underlying radial gradient. We quantify this using the metallicity deviation, which we defined as $\rm \Delta O/H=(12+log(O/H))_{obs} - (12+log(O/H))_{exp}$. Each \HII region's $\rm \Delta O/H$ is shown in the middle panel of Figure \ref{fig:Vrad_deltaOH_A1}. Within the optical disk, \HII regions exhibit small deviations from the expected metallicity, with a residual standard error of 0.049, consistent with a relatively well-mixed ISM. In contrast, the XUV disk regions exhibit a significantly larger scatter, with a residual standard error of 0.089, nearly twice that of the inner disk. The ALM regions, defined by $\rm \Delta O/H < -0.1$, show the largest deviations and drive this increase in scatter, indicating that the outer disk is being perturbed by the introduction of chemically distinct gas. 

Coupled with the measurement of the inwards-moving neutral gas in the outer disk, we consider this evidence for the flow of low metallicity gas from the extended disk, diluting existing ISM gas and triggering star formation (detected by young \HII regions) in the XUV disk of NGC~3344. This is one of the clearest observational links between radial gas flows and ALM \HII regions in the outer disk of a star-forming galaxy

\subsection{Inferring an Accretion Rate via HI Asymmetry}
If radial inflows are driven by external accretion, we expect the \HI disk to exhibit signatures of asymmetry, such as kinematic and density lopsidedness. Differences between the approaching and receding side rotation curves are an indication of kinematic lopsidedness. We measure the rotation curve in the approaching and receding side of the \HI disk using $\rm ^{3D}BAROLO$ and find that the two rotation curves differ from 8 to 13 kpc, with the maximum rotational velocity difference being $\sim$20 $\rm km~s^{-1}$.

Asymmetries within the \HI disk due to a density lopsidedness can be quantified using the normalized m=1 Fourier amplitude, which is defined to be ${\rm A1~(R)} = |\Sigma_j~I_j~e^{-i{\phi}_j}|/ \Sigma_j~I_j$, where $I_j$ and $\phi_j$ are the \HI intensity and azimuthal angle, respectively, at each pixel \textit{j} within each ring at galactocentric radius R \citep{Baldwin_1980MNRAS.193..313B, Rix_1995ApJ...447...82R,Zaritsky_1997ApJ...477..118Z, Saha_2014MNRAS.444..352S}. The radial profile of A1 is shown by the pink points in the bottom panel of Figure \ref{fig:Vrad_deltaOH_A1}. When $\rm A1>0.3$, the galaxy is considered strongly lopsided. In NGC~3344, A1 begins to increase at the start of the XUV disk, with the values reaching above 0.3 at $\sim$11 kpc. Following \cite{Zaritsky_1997ApJ...477..118Z}, we use the observed \HI lopsidedness to place a model-dependent upper limit on the global gas accretion rate, $\rm \dot{M}_{acc} = (A1\times M_{HI})/t_{lop}$, where $\rm t_{lop}$ is the characteristic lifetime of the lopsided distortion. Adopting $\rm A1_{max} = 0.64$ and  $\rm t_{lop} = 1~Gyr$ \citep{Rix_1995ApJ...447...82R} gives $\rm \dot{M}_{acc}$ of 1.76 $\rm M_\odot~yr^{-1}$. Because this value is an upper limit based on the maximum measured asymmetry, assigning it a conventional symmetric statistical uncertainty would not be meaningful. To illustrate the sensitivity of the estimate to the adopted asymmetry amplitude, we also calculate a characteristic value using the median $\rm A1$=0.10 between 6 and 11 kpc, before the sharp rise in $\rm A1$ in the outermost disk. This gives $\rm \dot{M}_{acc}$ of 0.27 $\rm M_\odot~yr^{-1}$. Thus, plausible choices of the asymmetry amplitude yield accretion-rate estimates of approximately 0.27–1.76 $\rm M_\odot~yr^{-1}$, with the upper end representing the formal upper-limit estimate. The dominant uncertainty is systematic and arises from associating the observed lopsidedness with accretion and from the assumed lifetime of the distortion. These estimates are consistent with a scenario in which accretion contributes to the strongly lopsided outer H I disk, after which the gas moves inward and may help sustain star formation and dilute the existing ISM.

\subsection{Insights from Chemical Evolution Models}\label{sec:chem_evol}

\begin{figure*}[t]
\centering
    \includegraphics[width = \linewidth]{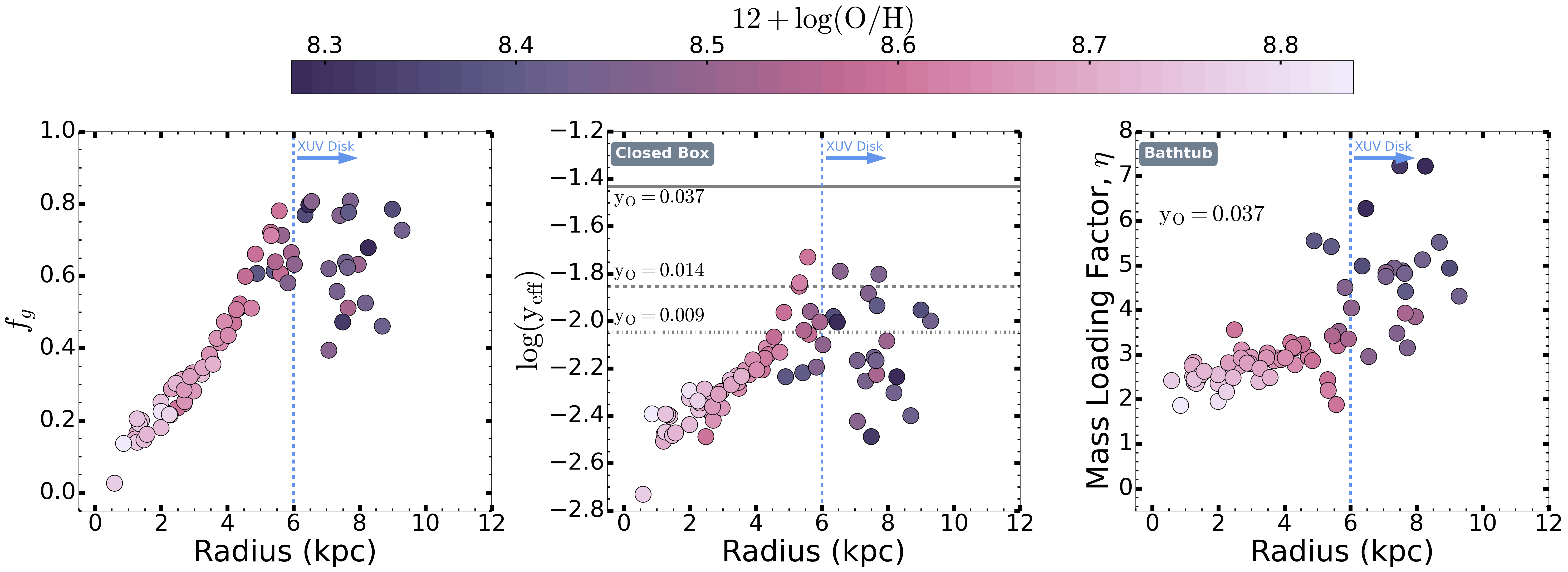}
\caption{The radial profile of the gas fraction, effective yield, and mass-loading factor is shown in the left, middle, and right panels, respectively. The color of the points represents the gas-phase metallicity of each region. The vertical dashed line at 6 kpc in each panel marks the onset of the XUV disk. The linear trend seen in all three radial profiles breaks down at this radius.}
\label{fig:radial_trends}
\end{figure*}

Chemical evolution models are a useful way to combine galaxy observables, such as metallicity, stellar mass, and gas mass, to test the level of chemical enrichment from star formation and the effects of outflows and inflows on the system's evolution \citep{Tinsley_1973ApJ...186...35T,Maeder_1992A&A...264..105M, Bouche_2010ApJ...718.1001B, Lilly_2013ApJ...772..119L}. Assuming the galaxy is not pre-enriched with metals and that no gas is able to enter or leave the system, we model the \HII regions of NGC 3344 as a closed box, which can be described by

\begin{equation}\label{equ:y_eff}
    {\rm{y_{eff}}} \equiv \frac{Z_{gas}}{{\rm{ln}}(1/f_{gas})}
\end{equation}

\noindent where $\rm Z_{gas}$ is the metallicity by mass and y$\rm _{eff}$ is the effective yield \cite{vandenBergh_1958AJ.....63..492V,Talbot_1971ApJ...170..409T}. The gas fraction, $f_{gas}$, is defined to be $f_{gas} \equiv {\rm M_{gas}/(M_{gas} + M_{\star}})$, where $\rm M_{gas}$ and $\rm M_{\star}$ are gas mass and stellar mass, respectively. For a closed box model, $\rm y_{eff}=y_O$, where $\rm y_O$ is the ``true'' nucleosynthetic yield of oxygen. Any deviation of $\rm y_{eff}$ from $\rm y_O$ is an indication of a departure from the closed box model (e.g. gas flows in and out of a system; \citealt{Edmunds_1990MNRAS.246..678E}).

The variation of measured $f_{gas}$ and calculated $\rm y_{eff}$ as a function of galactocentric radius is shown in the left and middle panels of Figure \ref{fig:radial_trends}. These values and those used to calculate them are given in Table \ref{tab:properties}. The points are color-coded by their gas-phase metallicity. Other ``true'' yield values reported in the literature \citep{Vincenzo_Yields_2016MNRAS.455.4183V,Barrera-Ballesteros_2018ApJ...852...74B} are shown as dotted and dashed lines. From 0 to 6 kpc, both $f_{gas}$ and y$\rm _{eff}$ increase with galactocentric radius, which is a telltale sign of a deviation from the closed box model. Both metal-rich outflows and metal-poor inflows can cause this. At the start of the XUV disk, this trend breaks down, and the scatter in the range of values increases. The deposition of low-metallicity gas carried by the radial flows likely breaks the closed-box assumption. 

The bathtub model of chemical evolution accounts for gas flowing in and out of the galaxy by introducing the mass-loading factor, $\eta = \rm \dot{M}_{outflow}/SFR$, and assuming the galaxy to be in steady-state (e.g the star-formation rate is equal to the gas inflow rate; \citealt{Lilly_2013ApJ...772..119L}). Following \cite{Lilly_2013ApJ...772..119L} and assuming the initial metallicity, Z$_0$, to be null, the return fraction, R, to be 0.455, and the yield, y, to be 0.037, we derive the radial profile of $\eta$ as shown in the right panel of Figure \ref{fig:radial_trends}. It reveals that the mass loading factor slowly increased out to about 6 kpc. Within the XUV disk, $\eta$ ranges from 3 to as high as 7.2. Regions with $\eta > 3$ in the XUV disk show unrealistic values of the mass-loading factor because there is little to no stars and thus no feedback mechanism able to push material out of the regions at the rates suggested by the $\eta$. The high values of $\eta$ are caused by the steady-state assumption breaking down in the outer disk where inflow rates are higher than the SFR. These inflows bring in metal-poor material and dilute the existing gas. In turn, following the inverse proportionality between $\eta$ and metallicity from Equ. 27a of \cite{Lilly_2013ApJ...772..119L}, the mass-loading factor is inflated to unrealistic values. The color of points in the right panel of Figure \ref{fig:radial_trends} shows that the regions with the highest $\eta$ in the XUV disk have the lowest metallicity.

We note that the scatter in the radial profiles of the XUV disk shown in Figure~\ref{fig:radial_trends} exhibits a pronounced spatial asymmetry. The regions with the largest deviations toward low gas fractions ($f_{\rm g}<0.6$), including several of the outermost regions, are concentrated in the southeastern quadrant of the galaxy—the portion of the disk that also appears lopsided in \HI. By contrast, the outermost regions in the northeastern quadrant generally have higher gas fractions and more closely follow the overall trend. The outer southeastern regions also have, on average, higher stellar masses than their northeastern counterparts. This spatial correspondence suggests that the observed scatter may be related not only to galactocentric radius but also to differences in the gas distribution and evolutionary history across the outer disk. The metallicities show no comparable quadrant-dependent trend, remaining low and exhibiting substantial scatter throughout the XUV disk. Although the origin of this asymmetry remains uncertain, spatially nonuniform gas accretion may influence the local gas supply and star-formation history.

\section{Conclusion} \label{sec:conclu}
We combine high-resolution VLA \HI 21cm imaging with MMT optical spectroscopy of 76 \HII regions to connect radial gas flows to star formation occurring in the XUV disk of the nearby galaxy, NGC 3344. The fluxes of the H$\alpha$ and [\ion{N}{2}] emission lines in our spectra are used to estimate the metallicity of the star-forming regions. Additionally, we model the extended \HI disk as concentric, tilted rings using $\rm ^{3D}BAROLO$ to measure radial velocities. Combining the metallicity information with the stellar and gas content, we perform chemical evolution modeling using the closed-box and bathtub models. Our main results and conclusions are:

\begin{enumerate}
    \item  The outer \HII regions (R$>$6 kpc) exhibit a scatter in metallicity almost twice that of inner regions. Several regions in the XUV disk have metallicities that are anomalously lower than those of other regions at the same radius.
    \item We measure radial velocities on the order of $\rm \sim6~km~s^{-1}$ from the start of the XUV disk (R = 6 kpc) moving inwards towards the galaxy's center.
    \item It is likely that the regions within the XUV disk of NGC 3344, with anomalously low metallicities, are caused by the neutral gas moving inwards and diluting the preexisting star-forming regions.  The inward-moving neutral gas is likely fueling the star formation occurring in NGC 3344's XUV disk.
    \item We quantify the lopsidedness of the \HI disk using the m=1 asymmetries and find $\rm A1>0.3$, hinting at an accretion event in the past $\sim$Gyr. We use the asymmetries to place an upper limit on the gas accretion rate and measure a global value of 1.76 $\rm M_\odot~yr^{-1}$.
    \item We find that the trends seen in the radial profile of gas fraction, effective yield, and mass-loading factor are disrupted at the start of the XUV disk due to gas flows.
\end{enumerate}

Previous studies have used ALM regions as indirect evidence for the recent accretion of low-metallicity gas, yet they have never identified the gas or measured its kinematics at the same location. This study, for the first time, links observed ALM regions to neutral gas moving radially inward. While we do not directly observe the gas diluting the star-forming regions, its kinematics reveal that the neutral gas reservoir in the outer disk is constantly replenished by gas moving towards the galactic center. This study illustrates how high-resolution imaging of gas can be combined with optical spectroscopy to reveal a connection between gas kinematics and chemical abundances in star-forming regions in the outskirts of spiral galaxies. 

\begin{acknowledgments}
We thank the reviewer for their constructive feedback which have made the paper more interesting and more likely to stimulate explanations for the presented findings.

This material is based upon work supported by the National Science Foundation Graduate
Research Fellowship Program under Grant No. 2233001. Any opinions, findings,
and conclusions or recommendations expressed in this material are those of the author(s)
and do not necessarily reflect the views of the National Science Foundation.

A.O. and S.B. acknowledge support from the NSF grants 2511242 and 2108159 as well as NASA ADAP grant 80NSSC21K0643.

E.D.T. was supported by the European Research Council (ERC) under grant agreement No. 101040751.

The Arizona State University authors acknowledge the twenty-three Native Nations that have inhabited this land for centuries. Arizona State University's four campuses are located in the Salt River Valley on ancestral territories of Indigenous peoples, including the Akimel O’odham (Pima) and Pee Posh (Maricopa) Indian Communities, whose care and keeping of these lands allow us to be here today.

We thank the support staff at the MMT Observatory, the Steward Observatory, the Vatican Advanced Technology Telescope, the Very Large Array, the National Radio Astronomy Observatory, and the Space Telescope Science Institute for their help with this project.

Observations reported here were obtained at the MMT Observatory, a joint facility of the Smithsonian Institution and the University of Arizona.

This work is also partly based on observations with the VATT: the Alice P. Lennon Telescope and the Thomas J. Bannan Astrophysics Facility.

The National Radio Astronomy Observatory is a facility of the National Science Foundation operated under cooperative agreement by Associated Universities, Inc.
 
This research made use of Photutils, an Astropy package for
detection and photometry of astronomical sources \citep{Bradley_larry_bradley_2023_7946442}.
\end{acknowledgments}




%
\facilities{HST, GALEX, MMT (Binospec), VATT, and VLA}

\software{$\rm ^{3D}BAROLO$ \citep{DiTeodoro_2015MNRAS.451.3021D},
          astropy \citep{astropy:2013,astropy:2018,astropy:2022},  
          \texttt{ipython/ jupyter} \citep{Perez_2007CSE.....9c..21P,Kluyver_2016ppap.book...87K}, 
          \texttt{matplotlib} \citep{Hunter:2007},
          \texttt{NumPy} \citep{harris2020array},
          photutils \citep{Bradley_larry_bradley_2023_7946442},
          \texttt{pPXF} \citep{Cappellari_2023MNRAS.526.3273C},
          \texttt{specutils} \citep{specutils}, and 
          \texttt{Python} from \url{https://www.python.org}
          }


\appendix

\section{BPT Diagram}\label{sec:BPT}

\begin{figure}[!h]
    \includegraphics[width=\linewidth]{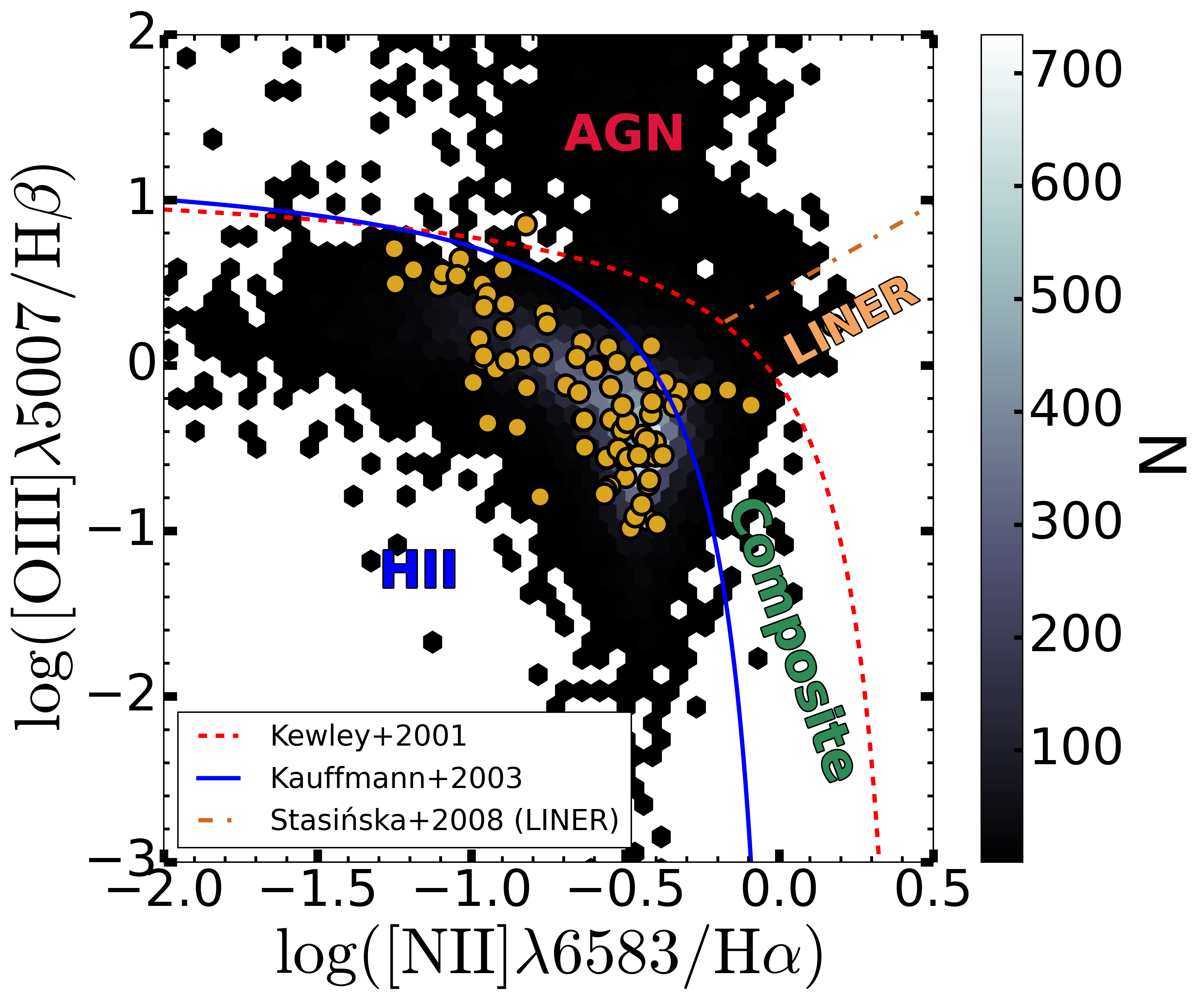}
\caption{``BPT'' Diagram for the 85 regions observed with the MMT/Binospec in comparison to the Calar Alto Legacy Integral Field Area(CALIFA) survey data \citep{EspinosaPonce_2020MNRAS.494.1622E}.}
\label{fig:BPT}
\end{figure}

We use the classic BPT diagram to classify our regions as star-forming or non-star-forming. Figure \ref{fig:BPT} shows where each region lies in BPT space.
\section{Errors on $\rm V_{rad}$}\label{vrad_err}
Errors on the kinematic parameters in $\rm ^{3D}BAROLO$ are estimated through a Monte Carlo exploration of the residual function around the best-fit solution in position-position-velocity space \citep{DiTeodoro_2015MNRAS.451.3021D,DiTeodoro_2021ApJ...923..220D}. The code samples the local parameter space and defines uncertainties as the range of parameter values corresponding to an arbitrary 5\% increase in the residual function. Consequently, these quantities should not be interpreted as formal statistical confidence intervals.

To assess the sensitivity of the radial velocity measurements, we also manually varied the radial velocity of individual rings by 1-5 $\rm km~s^{-1}$ and regenerated the model for each change. These tests showed that modest changes of only a few $\rm km~s^{-1}$ noticeably altered the residual maps, suggesting that the radial velocities are constrained at the level of several $\rm km~s^{-1}$, although the exact value depends on the ring and local parameter degeneracies.

Based on these tests, we adopt a representative uncertainty of approximately 5 $\rm km~s^{-1}$ for the radial velocities. This estimate is also consistent with the uncertainties reported for NGC 3344 by \cite{DiTeodoro_2021ApJ...923..220D}, who found typical radial velocity uncertainties of roughly 3 $\rm km~s^{-1}$ using an independent analysis.

The formal $\rm ^{3D}BAROLO$ uncertainty routine yields typical values of $-21$ and $+20$ $\rm km~s^{-1}$ for our model. However, these values primarily reflect the broad and highly degenerate residual surface (particularly the covariance between $\rm V_{\rm rad}$ and position angle) rather than the precision with which the radial velocities are constrained. In several rings the uncertainty estimation fails altogether because no well-defined residual minimum is identified. Therefore, we regard the formal BB uncertainties as indicative of parameter degeneracies rather than measurement precision.

Although the precise uncertainty on an individual ring is difficult to quantify, the inferred inflow is supported by the coherent behavior of multiple consecutive rings in the XUV disk, all of which exhibit negative radial velocities. It is therefore unlikely that the observed inflow signature arises solely from random uncertainties affecting individual rings.

\section{Table of Region Properties}\label{sec:tab_prop}

We provided all of our measured and calculated values from this work in Table \ref{tab:properties}.

\setlength{\tabcolsep}{2pt}
\begin{deluxetable*}{CCCCCCCCCCCCCCCCCCC}[h!]
\label{tab:properties}
\tablecaption{The positions and measured physical characteristics of the observed regions in NGC 3344 }
\tablewidth{0pt}
\tabletypesize{\tiny}
\tablehead{
\colhead{Region} & \colhead{Radius} & \colhead{R.A.} &  \colhead{Decl.}  & \colhead{F(H$\alpha$)} & \colhead{F(H$\alpha$)/F(H$\beta$)} &  \colhead{N2} & \colhead{$\rm 12+log(O/H)_{N2}$} & \colhead{$\rm log(M_\star)$} & \colhead{$\rm log(M_{HI})$} & \colhead{$\rm log(M_{H_2})$} & \colhead{$f_{gas}$} & \colhead{$\rm log(y_{eff})$} & \colhead{$\eta$} & \colhead{Region}\\
\colhead{Number} & \colhead{(kpc)} & \colhead{(degrees)} & \colhead{(degrees)} & \colhead{($\rm erg ~s^{-1} ~cm^{-2}$)} & \colhead{} & \colhead{} & \colhead{} & \colhead{($\rm M_\odot$)} & \colhead{($\rm M_\odot$)} & \colhead{($\rm M_\odot$)} & \colhead{} & \colhead{} & \colhead{for $\rm y_O = 0.037$} & \colhead{Type}
}
\startdata
1     & 7.5   & 160.8857 & 24.8708 & 85.02\pm5.11 & 3.44\pm0.25 & -1.189\pm0.043 & 8.32\pm0.024 & 7.46  & 7.29  &       & 0.473 & -2.49 & 7.23  & SF \\
2     & 7.3   & 160.8904 & 24.8727 & 28.14\pm1.10 & 4.21\pm0.32 & -0.947\pm0.074 & 8.44\pm0.038 & 7.37  & 7.34  &       & 0.557 & -2.25 & 4.94  & SF \\
3     & 8.2   & 160.9204 & 24.8795 & 116.87\pm5.60 & 3.58\pm0.28 & -0.961\pm0.045 & 8.44\pm0.023 & 7.45  & 7.37  &       & 0.525 & -2.30 & 5.13  & SF \\
4     & 8.7   & 160.9294 & 24.8826 & 32.70\pm1.11 & 3.37\pm0.23 & -0.995\pm0.040 & 8.42\pm0.020 & 7.49  & 7.29  &       & 0.461 & -2.40 & 5.52  & SF \\
5     & 7.7   & 160.9188 & 24.8831 & 23.92\pm0.99 & 3.88\pm0.45 & -0.773\pm0.047 & 8.53\pm0.024 & 7.46  & 7.35  &       & 0.511 & -2.22 & 3.93  & SF \\
6     & 5.9   & 160.8930 & 24.8832 & 36.71\pm1.55 & 4.14\pm0.20 & -0.762\pm0.042 & 8.54\pm0.022 & 7.18  & 7.35  &       & 0.665 & -2.00 & 3.35  & SF \\
7     & 6.0   & 160.9004 & 24.8852 & 51.52\pm1.86 & 4.09\pm0.19 & -0.852\pm0.027 & 8.49\pm0.013 & 7.23  & 7.34  &       & 0.632 & -2.10 & 4.05  & SF \\
8     & 5.5   & 160.8724 & 24.8852 & 38.69\pm1.63 & 4.69\pm0.32 & -0.753\pm0.039 & 8.54\pm0.020 & 7.22  & 7.34  &       & 0.639 & -2.04 & 3.42  & SF \\
9     & 5.8   & 160.8620 & 24.8852 & 37.11\pm1.31 & 4.61\pm0.17 & -0.894\pm0.033 & 8.47\pm0.016 & 7.30  & 7.31  &       & 0.581 & -2.19 & 4.50  & SF \\
10    & 7.7   & 160.8346 & 24.8882 & 97.87\pm6.78 & 3.87\pm0.32 & -1.036\pm0.048 & 8.40\pm0.025 & 7.00  & 7.41  &       & 0.776 & -1.93 & 4.41  & SF \\
11    & 7.1   & 160.8395 & 24.8894 & 108.75\pm6.19 & 4.49\pm0.32 & -0.949\pm0.029 & 8.44\pm0.015 & 7.23  & 7.32  &       & 0.620 & -2.17 & 4.75  & SF \\
12    & 7.1   & 160.9199 & 24.8898 & 121.86\pm3.56 & 3.65\pm0.23 & -0.885\pm0.023 & 8.48\pm0.012 & 7.64  & 7.33  &       & 0.394 & -2.42 & 4.84  & SF \\
13    & 4.7   & 160.8921 & 24.8917 & 335.40\pm14.63 & 4.05\pm0.19 & -0.595\pm0.026 & 8.63\pm0.015 & 7.47  & 7.36  & 5.32  & 0.511 & -2.13 & 2.95  & SF \\
14    & 4.9   & 160.8620 & 24.8924 & 232.89\pm11.41 & 3.98\pm0.20 & -1.047\pm0.029 & 8.39\pm0.015 & 7.29  & 7.35  &       & 0.607 & -2.23 & 5.55  & SF \\
15    & 4.3   & 160.8739 & 24.8933 & 47.18\pm2.07 & 3.62\pm0.21 & -0.633\pm0.031 & 8.60\pm0.017 & 7.43  & 7.31  &       & 0.507 & -2.16 & 3.16  & SF \\
16    & 4.5   & 160.8610 & 24.8958 & 208.52\pm18.83 & 4.22\pm0.45 & -0.694\pm0.051 & 8.57\pm0.027 & 7.38  & 7.38  & 6.39  & 0.599 & -2.07 & 3.23  & SF \\
17    & 3.7   & 160.8697 & 24.8985 & 71.24\pm3.81 & 3.52\pm0.31 & -0.548\pm0.033 & 8.65\pm0.020 & 7.58  & 7.33  &       & 0.428 & -2.21 & 2.86  & SF \\
18    & 3.5   & 160.8864 & 24.8991 & 116.02\pm5.41 & 4.22\pm0.25 & -0.565\pm0.025 & 8.64\pm0.015 & 7.68  & 7.32  &       & 0.371 & -2.28 & 3.04  & SF \\
19    & 3.5   & 160.8711 & 24.8998 & 75.69\pm2.23 & 4.41\pm0.29 & -0.547\pm0.019 & 8.65\pm0.012 & 7.65  & 7.31  &       & 0.383 & -2.26 & 2.93  & SF \\
20    & 3.8   & 160.8960 & 24.9009 & 12.84\pm1.24 & 3.50\pm0.55 & -0.415\pm0.048 &       & 7.57  & 7.27  & 7.02  & 0.511 &       &       & Non-SF \\
21    & 5.6   & 160.9157 & 24.9015 & 22.85\pm1.58 & 4.72\pm0.46 & -0.693\pm0.055 & 8.57\pm0.029 & 7.31  & 7.37  &       & 0.607 & -2.06 & 3.20  & SF \\
22    & 3.9   & 160.8594 & 24.9024 & 271.85\pm11.80 & 3.75\pm0.32 & -0.569\pm0.033 & 8.64\pm0.019 & 7.51  & 7.34  &       & 0.473 & -2.16 & 2.89  & SF \\
23    & 3.0   & 160.8881 & 24.9033 & 54.13\pm3.33 & 4.56\pm0.32 & -0.526\pm0.038 & 8.67\pm0.023 & 7.79  & 7.26  &       & 0.282 & -2.37 & 2.94  & SF \\
24    & 5.4   & 160.9155 & 24.9038 & 832.07\pm54.07 & 3.62\pm0.25 & -1.034\pm0.047 & 8.40\pm0.024 & 7.33  & 7.41  &       & 0.614 & -2.22 & 5.42  & SF \\
25    & 4.2   & 160.9048 & 24.9046 & 83.67\pm3.22 & 2.92\pm0.26 & -0.640\pm0.036 & 8.60\pm0.019 & 7.57  & 7.39  &       & 0.471 & -2.21 & 3.27  & SF \\
26    & 3.0   & 160.8683 & 24.9047 & 39.67\pm2.55 & 5.30\pm0.60 & -0.510\pm0.032 & 8.68\pm0.020 & 7.83  & 7.40  &       & 0.332 & -2.30 & 2.80  & SF \\
27    & 2.7   & 160.8864 & 24.9048 & 52.91\pm2.69 & 3.79\pm0.33 & -0.552\pm0.026 & 8.65\pm0.016 & 7.88  & 7.27  &       & 0.252 & -2.42 & 3.10  & SF \\
28    & 7.4   & 160.9327 & 24.9049 & 12.45\pm0.41 & 3.30\pm0.46 &       &       & 7.54  & 7.45  &       & 0.523 &       &       & SF \\
29    & 2.6   & 160.8829 & 24.9049 & 29.57\pm1.66 & 4.44\pm0.34 & -0.500\pm0.044 & 8.68\pm0.029 & 7.94  & 7.32  & 6.90  & 0.308 & -2.32 & 2.77  & SF \\
30    & 2.4   & 160.8816 & 24.9056 & 137.34\pm8.88 & 4.32\pm0.33 & -0.445\pm0.038 & 8.72\pm0.028 & 7.98  & 7.34  & 6.95  & 0.303 & -2.29 & 2.48  & SF \\
31    & 2.6   & 160.8690 & 24.9070 & 495.29\pm43.78 & 3.87\pm0.37 & -0.495\pm0.042 & 8.69\pm0.028 & 7.93  & 7.40  & 6.58  & 0.313 & -2.31 & 2.74  & SF \\
32    & 2.3   & 160.8706 & 24.9087 & 53.61\pm1.39 & 4.44\pm0.24 & -0.503\pm0.025 & 8.68\pm0.016 & 7.97  & 7.36  & 6.71  & 0.287 & -2.35 & 2.82  & SF \\
33    & 2.0   & 160.8769 & 24.9089 & 28.88\pm1.52 & 3.84\pm0.33 & -0.354\pm0.031 & 8.81 & 8.05  & 7.36  & 6.25  & 0.225 & -2.29 & 1.95  & SF \\
34    & 1.7   & 160.8737 & 24.9116 & 26.77\pm1.43 & 4.29\pm0.40 & -0.325\pm0.029 &       & 8.14  & 7.35  & 6.07  & 0.187 &       &       & Non-SF \\
35    & 1.5   & 160.8776 & 24.9126 & 15.21\pm1.46 & 5.49\pm1.02 & -0.250\pm0.066 &       & 8.23  & 7.34  & 7.09  & 0.213 &       &       & Non-SF \\
36    & 1.3   & 160.8749 & 24.9143 & 38.63\pm1.90 & 3.94\pm0.25 & -0.405\pm0.034 & 8.75\pm0.031 & 8.26  & 7.32  & 7.06  & 0.192 & -2.39 & 2.35  & SF \\
37    & 1.5   & 160.8715 & 24.9153 & 136.90\pm5.83 & 3.74\pm0.25 & -0.431\pm0.035 & 8.73\pm0.028 & 8.21  & 7.29  & 6.10  & 0.146 & -2.48 & 2.55  & SF \\
38    & 1.3   & 160.8730 & 24.9159 & 64.72\pm1.94 & 3.84\pm0.23 & -0.402\pm0.026 & 8.76\pm0.024 & 8.29  & 7.27  & 6.70  & 0.140 & -2.47 & 2.38  & SF \\
39    & 1.1   & 160.8742 & 24.9170 & 56.65\pm1.41 & 4.13\pm0.30 & -0.168\pm0.026 &       & 8.40  & 7.20  & 6.92  & 0.116 &       &       & Non-SF \\
40    & 0.8   & 160.8791 & 24.9170 & 20.71\pm0.65 & 3.23\pm0.39 & -0.092\pm0.026 &       & 8.57  & 7.02  & 7.34  & 0.104 &       &       & Non-SF \\
41    & 1.2   & 160.8718 & 24.9179 & 57.68\pm2.73 & 3.69\pm0.24 & -0.468\pm0.042 & 8.70\pm0.030 & 8.32  & 7.23  & 7.00  & 0.148 & -2.51 & 2.75  & SF \\
42    & 0.7   & 160.8827 & 24.9188 & 27.48\pm0.40 & 5.99\pm0.80 & 0.173\pm0.019 &       & 8.67  & 6.97  & 6.93  & 0.049 &       &       & Non-SF \\
43    & 1.2   & 160.8706 & 24.9198 & 32.36\pm1.00 & 3.30\pm0.37 & -0.424\pm0.019 & 8.74\pm0.015 & 8.32  & 7.21  & 7.01  & 0.146 & -2.48 & 2.51  & SF \\
44    & 1.3   & 160.8699 & 24.9218 & 196.59\pm7.73 & 3.68\pm0.23 & -0.484\pm0.032 & 8.69\pm0.021 & 8.30  & 7.19  & 7.14  & 0.164 & -2.49 & 2.83  & SF \\
45    & 0.1   & 160.8799 & 24.9223 & 1575.25\pm47.55 & 5.06\pm0.25 & -0.372\pm0.018 &       & 9.09  & 6.64  &       & 0.005 &       &       & Non-SF \\
46    & 1.4   & 160.8689 & 24.9239 & 45.62\pm2.19 & 4.39\pm0.34 & -0.426\pm0.030 & 8.74\pm0.024 & 8.26  & 7.22  & 7.24  & 0.200 & -2.40 & 2.47  & SF \\
47    & 0.6   & 160.8836 & 24.9241 & 45.98\pm2.11 & 3.87\pm0.28 & -0.397\pm0.026 & 8.76\pm0.025 & 8.79  & 6.99  & 6.39  & 0.026 & -2.73 & 2.42  & SF \\
48    & 2.0   & 160.8648 & 24.9262 & 35.52\pm1.41 & 6.07\pm0.66 & -0.413\pm0.025 & 8.75\pm0.021 & 8.11  & 7.32  & 7.03  & 0.251 & -2.33 & 2.35  & SF \\
49    & 2.3   & 160.8631 & 24.9276 & 23.90\pm1.35 & 3.89\pm0.42 & -0.418\pm0.030 & 8.74\pm0.026 & 8.01  & 7.32  & 4.69  & 0.215 & -2.37 & 2.41  & SF \\
\enddata
    
\end{deluxetable*}

\begin{deluxetable*}{CCCCCCCCCCCCCCCCCCC}[h!]
\tablewidth{0pt}
\tabletypesize{\tiny}
\tablehead{
\colhead{Region} & \colhead{Radius} & \colhead{R.A.} &  \colhead{Decl.}  & \colhead{F(H$\alpha$)} & \colhead{F(H$\alpha$)/F(H$\beta$)} &  \colhead{N2} & \colhead{$\rm 12+log(O/H)_{N2}$} & \colhead{$\rm log(M_\star)$} & \colhead{$\rm log(M_{HI})$} & \colhead{$\rm log(M_{H_2})$} & \colhead{$f_{gas}$} & \colhead{$\rm log(y_{eff})$} & \colhead{$\eta$} & \colhead{Region}\\
\colhead{Number} & \colhead{(kpc)} & \colhead{(degrees)} & \colhead{(degrees)} & \colhead{($\rm erg ~s^{-1} ~cm^{-2}$)} & \colhead{} & \colhead{} & \colhead{} & \colhead{($\rm M_\odot$)} & \colhead{($\rm M_\odot$)} & \colhead{($\rm M_\odot$)} & \colhead{} & \colhead{} & \colhead{for $\rm y_O = 0.037$} & \colhead{Type}
}
\startdata
50    & 0.9   & 160.8823 & 24.9279 & 34.66\pm1.98 & 6.66\pm0.45 & -0.340\pm0.043 & 8.84 & 8.54  & 7.19  & 7.41  & 0.136 & -2.39 & 1.86  & SF \\
51    & 2.5   & 160.8621 & 24.9296 & 244.78\pm12.09 & 3.55\pm0.24 & -0.634\pm0.026 & 8.60\pm0.014 & 7.94  & 7.29  &       & 0.234 & -2.49 & 3.56  & SF \\
52    & 1.3   & 160.8841 & 24.9303 & 393.47\pm37.01 & 3.28\pm0.40 & -0.422\pm0.057 & 8.74\pm0.048 & 8.33  & 7.29  & 7.32  & 0.205 & -2.39 & 2.45  & SF \\
53    & 1.6   & 160.8861 & 24.9316 & 125.68\pm5.38 & 4.28\pm0.28 & -0.447\pm0.027 & 8.72\pm0.020 & 8.22  & 7.29  & 6.64  & 0.162 & -2.47 & 2.63  & SF \\
54    & 2.7   & 160.8622 & 24.9326 & 45.49\pm1.74 & 3.82\pm0.28 & -0.505\pm0.033 & 8.68\pm0.021 & 7.91  & 7.29  & 4.47  & 0.246 & -2.40 & 2.86  & SF \\
55    & 3.2   & 160.8585 & 24.9343 & 13.40\pm0.71 & 5.18\pm0.89 & -0.435\pm0.036 & 8.73\pm0.029 & 7.76  & 7.32  &       & 0.328 & -2.25 & 2.39  & SF \\
56    & 2.0   & 160.8855 & 24.9350 & 19.88\pm0.81 & 2.53\pm0.19 & -0.341\pm0.034 &       & 8.05  & 7.28  &       & 0.184 &       &       & Non-SF \\
57    & 2.2   & 160.8882 & 24.9359 & 24.76\pm1.05 & 4.47\pm0.38 & -0.379\pm0.039 & 8.78\pm0.046 & 7.96  & 7.27  &       & 0.217 & -2.34 & 2.16  & SF \\
58    & 2.0   & 160.8830 & 24.9360 & 55.72\pm2.49 & 4.61\pm0.26 & -0.436\pm0.024 & 8.73\pm0.019 & 8.05  & 7.26  &       & 0.180 & -2.44 & 2.55  & SF \\
59    & 2.7   & 160.8936 & 24.9361 & 29.75\pm1.81 & 4.68\pm0.56 & -0.523\pm0.047 & 8.67\pm0.029 & 7.86  & 7.33  &       & 0.285 & -2.36 & 2.92  & SF \\
60    & 3.4   & 160.8583 & 24.9366 & 48.68\pm1.96 & 4.23\pm0.33 & -0.457\pm0.039 & 8.71\pm0.028 & 7.75  & 7.37  &       & 0.357 & -2.23 & 2.48  & SF \\
61    & 4.4   & 160.8498 & 24.9366 & 59.52\pm2.82 & 4.12\pm0.27 & -0.641\pm0.037 & 8.60\pm0.020 & 7.40  & 7.31  &       & 0.523 & -2.14 & 3.16  & SF \\
62    & 3.8   & 160.8567 & 24.9388 & 36.16\pm1.59 & 4.64\pm0.46 & -0.556\pm0.024 & 8.65\pm0.014 & 7.66  & 7.38  &       & 0.415 & -2.23 & 2.92  & SF \\
63    & 2.9   & 160.8905 & 24.9398 & 187.50\pm7.64 & 3.45\pm0.30 & -0.510\pm0.023 & 8.68\pm0.015 & 7.88  & 7.36  & 6.55  & 0.321 & -2.31 & 2.81  & SF \\
64    & 3.3   & 160.8864 & 24.9442 & 366.10\pm25.47 & 4.28\pm0.31 & -0.493\pm0.038 & 8.69\pm0.025 & 7.86  & 7.44  & 5.73  & 0.346 & -2.27 & 2.69  & SF \\
65    & 3.6   & 160.8915 & 24.9447 & 68.85\pm3.81 & 3.99\pm0.32 & -0.457\pm0.028 & 8.71\pm0.020 & 7.74  & 7.35  & 5.06  & 0.356 & -2.23 & 2.49  & SF \\
66    & 7.4   & 160.8284 & 24.9447 & 60.95\pm2.35 & 3.87\pm0.22 & -0.897\pm0.030 & 8.47\pm0.015 & 6.96  & 7.35  &       & 0.768 & -1.88 & 3.48  & SF \\
67    & 7.7   & 160.8266 & 24.9464 & 83.66\pm4.02 & 3.63\pm0.18 & -0.921\pm0.022 & 8.46\pm0.011 & 6.91  & 7.41  &       & 0.808 & -1.80 & 3.15  & SF \\
68    & 4.3   & 160.8608 & 24.9472 & 24.20\pm0.88 & 4.40\pm0.33 & -0.561\pm0.022 & 8.65\pm0.013 & 7.43  & 7.31  &       & 0.510 & -2.11 & 2.77  & SF \\
69    & 4.0   & 160.8914 & 24.9483 & 26.67\pm0.71 & 4.88\pm0.32 & -0.563\pm0.016 & 8.64\pm0.009 & 7.49  & 7.25  &       & 0.435 & -2.21 & 2.93  & SF \\
70    & 6.1   & 160.8440 & 24.9501 & 31.44\pm1.30 & 3.62\pm0.33 & -0.824\pm0.052 &       & 7.25  & 7.21  &       & 0.552 &       &       & Non-SF \\
71    & 5.6   & 160.8544 & 24.9544 & 79.38\pm2.21 & 3.46\pm0.19 & -0.835\pm0.030 & 8.50\pm0.015 & 7.00  & 7.26  &       & 0.713 & -1.96 & 3.52  & SF \\
72    & 4.8   & 160.8907 & 24.9546 & 22.42\pm0.80 & 3.27\pm0.17 & -0.672\pm0.041 & 8.58\pm0.022 & 7.27  & 7.43  & 5.14  & 0.661 & -1.96 & 2.86  & SF \\
73    & 5.6   & 160.8968 & 24.9578 & 30.12\pm1.36 & 4.36\pm0.35 & -0.651\pm0.036 & 8.60\pm0.019 & 7.09  & 7.40  & 6.87  & 0.780 & -1.73 & 1.88  & SF \\
74    & 9.0   & 160.8231 & 24.9583 & 22.08\pm0.61 & 2.89\pm0.31 & -1.108\pm0.075 & 8.36\pm0.040 & 6.92  & 7.35  &       & 0.785 & -1.95 & 4.94  & SF \\
75    & 5.3   & 160.8894 & 24.9584 & 114.20\pm5.87 & 2.83\pm0.23 & -0.601\pm0.023 & 8.62\pm0.013 & 7.23  & 7.50  & 5.14  & 0.712 & -1.84 & 2.20  & SF \\
76    & 7.6   & 160.8375 & 24.9588 & 21.06\pm1.45 & 4.30\pm0.49 & -0.976\pm0.040 & 8.43\pm0.020 & 7.09  & 7.21  &       & 0.638 & -2.15 & 4.88  & SF \\
77    & 5.3   & 160.8766 & 24.9592 & 38.33\pm0.92 & 3.25\pm0.28 & -0.657\pm0.029 & 8.59\pm0.016 & 7.16  & 7.44  &       & 0.721 & -1.85 & 2.44  & SF \\
78    & 6.6   & 160.8997 & 24.9641 & 107.04\pm3.39 & 3.08\pm0.18 & -0.889\pm0.023 & 8.47\pm0.012 & 7.00  & 7.50  &       & 0.806 & -1.79 & 2.96  & SF \\
79    & 8.3   & 160.8364 & 24.9646 & 131.93\pm5.60 & 3.40\pm0.29 & -1.247\pm0.030 & 8.28\pm0.017 & 7.10  & 7.30  &       & 0.678 & -2.23 & 7.22  & SF \\
80    & 8.0   & 160.8403 & 24.9648 & 34.69\pm1.53 & 3.48\pm0.39 & -0.821\pm0.055 & 8.51\pm0.028 & 7.16  & 7.27  &       & 0.632 & -2.08 & 3.85  & SF \\
81    & 9.3   & 160.8271 & 24.9661 & 25.65\pm0.87 & 3.52\pm0.18 & -0.965\pm0.066 & 8.43\pm0.034 & 7.01  & 7.31  &       & 0.727 & -2.00 & 4.31  & SF \\
82    & 6.4   & 160.8819 & 24.9665 & 1266.04\pm89.71 & 3.67\pm0.32 & -1.095\pm0.038 & 8.37\pm0.020 & 7.10  & 7.50  &       & 0.769 & -1.98 & 4.99  & SF \\
83    & 6.5   & 160.8843 & 24.9672 & 1522.56\pm59.81 & 3.66\pm0.24 & -1.251\pm0.024 & 8.28\pm0.014 & 7.08  & 7.54  &       & 0.797 & -2.00 & 6.28  & SF \\
84    & 7.6   & 160.9057 & 24.9699 & 443.87\pm20.59 & 3.75\pm0.21 & -0.960\pm0.024 & 8.44\pm0.012 & 7.22  & 7.31  &       & 0.624 & -2.17 & 4.82  & SF \\
85    & 8.9   & 160.8846 & 24.9838 & 21.74\pm0.93 & 4.67\pm0.34 & -0.777\pm0.052 & 8.53\pm0.027 &  & 7.47  &       &  &  &  & SF \\
\enddata
\tablenotetext{}{\scriptsize \tablecomments{The listed H$\alpha$ fluxes are the observed spectroscopic fluxes and have not been corrected for internal dust attenuation. Both the H$\alpha$ fluxes and their uncertainties are given in units of $10^{-16}\ {\rm erg\ s^{-1}\ cm^{-2}}$. Blank entries in the N2 column indicate regions in which the \ion{N}{2} $\lambda6583$ emission line was not detected. Blank entries in the $12+\log({\rm O/H})_{\rm N2}$ column indicate either that N2 was unavailable or that the N2 metallicity calibration was not valid for that measurement. Because the calculations of $y_{\rm eff}$ and $\eta$ require a valid gas-phase metallicity, the corresponding entries in those columns are also left blank.}}
\end{deluxetable*}

\bibliography{sample7}{}

@ARTICLE{Bertin_1996A&AS..117..393B,
       author = {{Bertin}, E. and {Arnouts}, S.},
        title = "{SExtractor: Software for source extraction.}",
      journal = {\aaps},
         year = "1996",
        month = "Jun",
       volume = {117},
        pages = {393-404},
          doi = {10.1051/aas:1996164},
       adsurl = {https://ui.adsabs.harvard.edu/abs/1996A&AS..117..393B}
}

@ARTICLE{Padave_2025ApJ...986..145P,
       author = {{Padave}, Mansi and {Borthakur}, Sanchayeeta and {Jansen}, Rolf A. and {Thilker}, David A. and {Monkiewicz}, Jacqueline and {Windhorst}, Rogier A.},
        title = "{DIISC-V: Variations in H{\ensuremath{\alpha}}-to-FUV Star Formation Rate Ratios Across Star-forming Regions in Nearby Galaxies}",
      journal = {\apj},
         year = 2025,
        month = jun,
       volume = {986},
       number = {2},
          eid = {145},
        pages = {145},
          doi = {10.3847/1538-4357/adce7b},
archivePrefix = {arXiv},
       eprint = {2407.16690},
 primaryClass = {astro-ph.GA},
       adsurl = {https://ui.adsabs.harvard.edu/abs/2025ApJ...986..145P}
}

@ARTICLE{kansky_2019PASP..131g5005K,
       author = {{Kansky}, Jan and {Chilingarian}, Igor and {Fabricant}, Daniel and {Matthews}, Anne and {Moran}, Sean and {Paegert}, Martin and {Duane Gibson}, J. and {Porter}, Dallan and {Roll}, John},
        title = "{Binospec Software System}",
      journal = {\pasp},
         year = 2019,
        month = jul,
       volume = {131},
       number = {1001},
        pages = {075005},
          doi = {10.1088/1538-3873/ab1ceb},
archivePrefix = {arXiv},
       eprint = {1905.03321},
 primaryClass = {astro-ph.IM},
       adsurl = {https://ui.adsabs.harvard.edu/abs/2019PASP..131g5005K}
}

@ARTICLE{Olvera_2024ApJ...976..205O,
       author = {{Olvera}, Alejandro J. and {Borthakur}, Sanchayeeta and {Padave}, Mansi and {Heckman}, Timothy and {Gim}, Hansung B. and {Koplitz}, Brad and {Dupuis}, Christopher and {Momjian}, Emmanuel and {Jansen}, Rolf A.},
        title = "{DIISC-IV. DIISCovery of Anomalously Low Metallicity H II Regions in NGC 99: Indirect Evidence of Gas Inflows}",
      journal = {\apj},
         year = 2024,
        month = dec,
       volume = {976},
       number = {2},
          eid = {205},
        pages = {205},
          doi = {10.3847/1538-4357/ad8238},
archivePrefix = {arXiv},
       eprint = {2408.08303},
 primaryClass = {astro-ph.GA},
       adsurl = {https://ui.adsabs.harvard.edu/abs/2024ApJ...976..205O}
}

@ARTICLE{Cappellari_2023MNRAS.526.3273C,
       author = {{Cappellari}, Michele},
        title = "{Full spectrum fitting with photometry in PPXF: stellar population versus dynamical masses, non-parametric star formation history and metallicity for 3200 LEGA-C galaxies at redshift z {\ensuremath{\approx}} 0.8}",
      journal = {\mnras},
         year = 2023,
        month = dec,
       volume = {526},
       number = {3},
        pages = {3273-3300},
          doi = {10.1093/mnras/stad2597},
archivePrefix = {arXiv},
       eprint = {2208.14974},
 primaryClass = {astro-ph.GA},
       adsurl = {https://ui.adsabs.harvard.edu/abs/2023MNRAS.526.3273C}
}

@ARTICLE{Searle_1971ApJ...168..327S,
       author = {{Searle}, Leonard},
        title = "{Evidence for Composition Gradients across the Disks of Spiral Galaxies}",
      journal = {\apj},
         year = 1971,
        month = sep,
       volume = {168},
        pages = {327},
          doi = {10.1086/151090},
       adsurl = {https://ui.adsabs.harvard.edu/abs/1971ApJ...168..327S}
}

@ARTICLE{Alloin_1979A&A....78..200A,
       author = {{Alloin}, D. and {Collin-Souffrin}, S. and {Joly}, M. and {Vigroux}, L.},
        title = "{Nitrogen and oxygen abundances in galaxies.}",
      journal = {\aap},
         year = 1979,
        month = sep,
       volume = {78},
        pages = {200-216},
       adsurl = {https://ui.adsabs.harvard.edu/abs/1979A&A....78..200A}
}

@ARTICLE{Kewley_2002ApJS..142...35K,
       author = {{Kewley}, L.~J. and {Dopita}, M.~A.},
        title = "{Using Strong Lines to Estimate Abundances in Extragalactic H II Regions and Starburst Galaxies}",
      journal = {\apjs},
         year = 2002,
        month = sep,
       volume = {142},
       number = {1},
        pages = {35-52},
          doi = {10.1086/341326},
archivePrefix = {arXiv},
       eprint = {astro-ph/0206495},
 primaryClass = {astro-ph},
       adsurl = {https://ui.adsabs.harvard.edu/abs/2002ApJS..142...35K}
}

@ARTICLE{Marino_2013A&A...559A.114M,
       author = {{Marino}, R.~A. and {Rosales-Ortega}, F.~F. and {S{\'a}nchez}, S.~F. and {Gil de Paz}, A. and {V{\'\i}lchez}, J. and {Miralles-Caballero}, D. and {Kehrig}, C. and {P{\'e}rez-Montero}, E. and {Stanishev}, V. and {Iglesias-P{\'a}ramo}, J. and {D{\'\i}az}, A.~I. and {Castillo-Morales}, A. and {Kennicutt}, R. and {L{\'o}pez-S{\'a}nchez}, A.~R. and {Galbany}, L. and {Garc{\'\i}a-Benito}, R. and {Mast}, D. and {Mendez-Abreu}, J. and {Monreal-Ibero}, A. and {Husemann}, B. and {Walcher}, C.~J. and {Garc{\'\i}a-Lorenzo}, B. and {Masegosa}, J. and {Del Olmo Orozco}, A. and {Mour{\~a}o}, A.~M. and {Ziegler}, B. and {Moll{\'a}}, M. and {Papaderos}, P. and {S{\'a}nchez-Bl{\'a}zquez}, P. and {Gonz{\'a}lez Delgado}, R.~M. and {Falc{\'o}n-Barroso}, J. and {Roth}, M.~M. and {van de Ven}, G. and {CALIFA Team}},
        title = "{The O3N2 and N2 abundance indicators revisited: improved calibrations based on CALIFA and T$_{e}$-based literature data}",
      journal = {\aap},
         year = 2013,
        month = nov,
       volume = {559},
          eid = {A114},
        pages = {A114},
          doi = {10.1051/0004-6361/201321956},
archivePrefix = {arXiv},
       eprint = {1307.5316},
 primaryClass = {astro-ph.CO},
       adsurl = {https://ui.adsabs.harvard.edu/abs/2013A&A...559A.114M}
}

@ARTICLE{Curti_FMR_2020MNRAS.491..944C,
       author = {{Curti}, Mirko and {Mannucci}, Filippo and {Cresci}, Giovanni and {Maiolino}, Roberto},
        title = "{The mass-metallicity and the fundamental metallicity relation revisited on a fully T$_{e}$-based abundance scale for galaxies}",
      journal = {\mnras},
         year = 2020,
        month = jan,
       volume = {491},
       number = {1},
        pages = {944-964},
          doi = {10.1093/mnras/stz2910},
archivePrefix = {arXiv},
       eprint = {1910.00597},
 primaryClass = {astro-ph.GA},
       adsurl = {https://ui.adsabs.harvard.edu/abs/2020MNRAS.491..944C}
}

@software{Newville_2021zndo....598352N,
       author = {{Newville}, Matthew and {Otten}, Renee and {Nelson}, Andrew and {Stensitzki}, Till and {Ingargiola}, Antonino and {Allan}, Daniel and {Fox}, Austin and {Carter}, Faustin and {Rawlik}, Michal},
        title = "{LMFIT: Non-Linear Least-Squares Minimization and Curve-Fitting for Python}",
         year = 2025,
        month = mar,
          eid = {10.5281/zenodo.598352},
          doi = {10.5281/zenodo.598352},
      version = {1.3.3},
    publisher = {Zenodo},
       adsurl = {https://ui.adsabs.harvard.edu/abs/2021zndo....598352N}
}

@ARTICLE{Thilker_2007ApJS..173..538T,
       author = {{Thilker}, David A. and {Bianchi}, Luciana and {Meurer}, Gerhardt and {Gil de Paz}, Armando and {Boissier}, Samuel and {Madore}, Barry F. and {Boselli}, Alessandro and {Ferguson}, Annette M.~N. and {Mu{\~n}oz-Mateos}, Juan Carlos and {Madsen}, Greg J. and {Hameed}, Salman and {Overzier}, Roderik A. and {Forster}, Karl and {Friedman}, Peter G. and {Martin}, D. Christopher and {Morrissey}, Patrick and {Neff}, Susan G. and {Schiminovich}, David and {Seibert}, Mark and {Small}, Todd and {Wyder}, Ted K. and {Donas}, Jos{\'e} and {Heckman}, Timothy M. and {Lee}, Young-Wook and {Milliard}, Bruno and {Rich}, R. Michael and {Szalay}, Alex S. and {Welsh}, Barry Y. and {Yi}, Sukyoung K.},
        title = "{A Search for Extended Ultraviolet Disk (XUV-Disk) Galaxies in the Local Universe}",
      journal = {\apjs},
         year = 2007,
        month = dec,
       volume = {173},
       number = {2},
        pages = {538-571},
          doi = {10.1086/523853},
archivePrefix = {arXiv},
       eprint = {0712.3555},
 primaryClass = {astro-ph},
       adsurl = {https://ui.adsabs.harvard.edu/abs/2007ApJS..173..538T}
}

@ARTICLE{GildePaz_2005ApJ...627L..29G,
       author = {{Gil de Paz}, A. and {Madore}, B.~F. and {Boissier}, S. and {Swaters}, R. and {Popescu}, C.~C. and {Tuffs}, R.~J. and {Sheth}, K. and {Kennicutt}, Jr., R.~C. and {Bianchi}, L. and {Thilker}, D. and {Martin}, D.~C.},
        title = "{Discovery of an Extended Ultraviolet Disk in the Nearby Galaxy NGC 4625}",
      journal = {\apjl},
         year = 2005,
        month = jul,
       volume = {627},
       number = {1},
        pages = {L29-L32},
          doi = {10.1086/432054},
archivePrefix = {arXiv},
       eprint = {astro-ph/0506357},
 primaryClass = {astro-ph},
       adsurl = {https://ui.adsabs.harvard.edu/abs/2005ApJ...627L..29G}
}

@ARTICLE{Padave_2021ApJ...923..199P,
       author = {{Padave}, Mansi and {Borthakur}, Sanchayeeta and {Gim}, Hansung B. and {Jansen}, Rolf A. and {Thilker}, David and {Heckman}, Timothy and {Kennicutt}, Robert C. and {Momjian}, Emmanuel and {Fox}, Andrew J.},
        title = "{DIISC-II: Unveiling the Connections between Star Formation and Interstellar Medium in the Extended Ultraviolet Disk of NGC 3344}",
      journal = {\apj},
         year = 2021,
        month = dec,
       volume = {923},
       number = {2},
          eid = {199},
        pages = {199},
          doi = {10.3847/1538-4357/ac2c01},
archivePrefix = {arXiv},
       eprint = {2110.07590},
 primaryClass = {astro-ph.GA},
       adsurl = {https://ui.adsabs.harvard.edu/abs/2021ApJ...923..199P}
}

@ARTICLE{Mihos_1996ApJ...464..641M,
       author = {{Mihos}, J. Christopher and {Hernquist}, Lars},
        title = "{Gasdynamics and Starbursts in Major Mergers}",
      journal = {\apj},
         year = 1996,
        month = jun,
       volume = {464},
        pages = {641},
          doi = {10.1086/177353},
archivePrefix = {arXiv},
       eprint = {astro-ph/9512099},
 primaryClass = {astro-ph},
       adsurl = {https://ui.adsabs.harvard.edu/abs/1996ApJ...464..641M}
}

@ARTICLE{Bush_2010ApJ...713..780B,
       author = {{Bush}, Stephanie J. and {Cox}, T.~J. and {Hayward}, Christopher C. and {Thilker}, David and {Hernquist}, Lars and {Besla}, Gurtina},
        title = "{Spiral-Induced Star Formation in the Outer Disks of Galaxies}",
      journal = {\apj},
         year = 2010,
        month = apr,
       volume = {713},
       number = {2},
        pages = {780-799},
          doi = {10.1088/0004-637X/713/2/780},
archivePrefix = {arXiv},
       eprint = {1003.5672},
 primaryClass = {astro-ph.GA},
       adsurl = {https://ui.adsabs.harvard.edu/abs/2010ApJ...713..780B}
}

@ARTICLE{Dekel_2023MNRAS.523.3201D,
       author = {{Dekel}, Avishai and {Sarkar}, Kartick C. and {Birnboim}, Yuval and {Mandelker}, Nir and {Li}, Zhaozhou},
        title = "{Efficient formation of massive galaxies at cosmic dawn by feedback-free starbursts}",
      journal = {\mnras},
         year = 2023,
        month = aug,
       volume = {523},
       number = {3},
        pages = {3201-3218},
          doi = {10.1093/mnras/stad1557},
archivePrefix = {arXiv},
       eprint = {2303.04827},
 primaryClass = {astro-ph.GA},
       adsurl = {https://ui.adsabs.harvard.edu/abs/2023MNRAS.523.3201D}
}

@ARTICLE{Bigiel_2008AJ....136.2846B,
       author = {{Bigiel}, F. and {Leroy}, A. and {Walter}, F. and {Brinks}, E. and {de Blok}, W.~J.~G. and {Madore}, B. and {Thornley}, M.~D.},
        title = "{The Star Formation Law in Nearby Galaxies on Sub-Kpc Scales}",
      journal = {\aj},
         year = 2008,
        month = dec,
       volume = {136},
       number = {6},
        pages = {2846-2871},
          doi = {10.1088/0004-6256/136/6/2846},
archivePrefix = {arXiv},
       eprint = {0810.2541},
 primaryClass = {astro-ph},
       adsurl = {https://ui.adsabs.harvard.edu/abs/2008AJ....136.2846B}
}

@ARTICLE{Keres_2005MNRAS.363....2K,
       author = {{Kere{\v{s}}}, Du{\v{s}}an and {Katz}, Neal and {Weinberg}, David H. and {Dav{\'e}}, Romeel},
        title = "{How do galaxies get their gas?}",
      journal = {\mnras},
         year = 2005,
        month = oct,
       volume = {363},
       number = {1},
        pages = {2-28},
          doi = {10.1111/j.1365-2966.2005.09451.x},
archivePrefix = {arXiv},
       eprint = {astro-ph/0407095},
 primaryClass = {astro-ph},
       adsurl = {https://ui.adsabs.harvard.edu/abs/2005MNRAS.363....2K}
}

@ARTICLE{Hafen_2022MNRAS.514.5056H,
       author = {{Hafen}, Zachary and {Stern}, Jonathan and {Bullock}, James and {Gurvich}, Alexander B. and {Yu}, Sijie and {Faucher-Gigu{\`e}re}, Claude-Andr{\'e} and {Fielding}, Drummond B. and {Angl{\'e}s-Alc{\'a}zar}, Daniel and {Quataert}, Eliot and {Wetzel}, Andrew and {Starkenburg}, Tjitske and {Boylan-Kolchin}, Michael and {Moreno}, Jorge and {Feldmann}, Robert and {El-Badry}, Kareem and {Chan}, T.~K. and {Trapp}, Cameron and {Kere{\v{s}}}, Du{\v{s}}an and {Hopkins}, Philip F.},
        title = "{Hot-mode accretion and the physics of thin-disc galaxy formation}",
      journal = {\mnras},
         year = 2022,
        month = aug,
       volume = {514},
       number = {4},
        pages = {5056-5073},
          doi = {10.1093/mnras/stac1603},
archivePrefix = {arXiv},
       eprint = {2201.07235},
 primaryClass = {astro-ph.GA},
       adsurl = {https://ui.adsabs.harvard.edu/abs/2022MNRAS.514.5056H}
}

@ARTICLE{Walter_2008AJ....136.2563W,
       author = {{Walter}, Fabian and {Brinks}, Elias and {de Blok}, W.~J.~G. and {Bigiel}, Frank and {Kennicutt}, Jr., Robert C. and {Thornley}, Michele D. and {Leroy}, Adam},
        title = "{THINGS: The H I Nearby Galaxy Survey}",
      journal = {\aj},
         year = 2008,
        month = dec,
       volume = {136},
       number = {6},
        pages = {2563-2647},
          doi = {10.1088/0004-6256/136/6/2563},
archivePrefix = {arXiv},
       eprint = {0810.2125},
 primaryClass = {astro-ph},
       adsurl = {https://ui.adsabs.harvard.edu/abs/2008AJ....136.2563W}
}

@ARTICLE{Schmidt_2016MNRAS.457.2642S,
       author = {{Schmidt}, Tobias M. and {Bigiel}, Frank and {Klessen}, Ralf S. and {de Blok}, W.~J.~G.},
        title = "{Radial gas motions in The H I Nearby Galaxy Survey (THINGS)}",
      journal = {\mnras},
         year = 2016,
        month = apr,
       volume = {457},
       number = {3},
        pages = {2642-2664},
          doi = {10.1093/mnras/stw011},
archivePrefix = {arXiv},
       eprint = {1601.01689},
 primaryClass = {astro-ph.GA},
       adsurl = {https://ui.adsabs.harvard.edu/abs/2016MNRAS.457.2642S}
}

@ARTICLE{DiTeodoro_2021ApJ...923..220D,
       author = {{Di Teodoro}, Enrico M. and {Peek}, J.~E.~G.},
        title = "{Radial Motions and Radial Gas Flows in Local Spiral Galaxies}",
      journal = {\apj},
         year = 2021,
        month = dec,
       volume = {923},
       number = {2},
          eid = {220},
        pages = {220},
          doi = {10.3847/1538-4357/ac2cbd},
archivePrefix = {arXiv},
       eprint = {2110.01618},
 primaryClass = {astro-ph.GA},
       adsurl = {https://ui.adsabs.harvard.edu/abs/2021ApJ...923..220D}
}

@ARTICLE{Hwang_2019ApJ...872..144H,
       author = {{Hwang}, Hsiang-Chih and {Barrera-Ballesteros}, Jorge K. and {Heckman}, Timothy M. and {Rowlands}, Kate and {Lin}, Lihwai and {Rodriguez-Gomez}, Vicente and {Pan}, Hsi-An and {Hsieh}, Bau-Ching and {S{\'a}nchez}, Sebastian and {Bizyaev}, Dmitry and {S{\'a}nchez Almeida}, Jorge and {Thilker}, David A. and {Lotz}, Jennifer M. and {Jones}, Amy and {Nair}, Preethi and {Andrews}, Brett H. and {Drory}, Niv},
        title = "{Anomalously Low-metallicity Regions in MaNGA Star-forming Galaxies: Accretion Caught in Action?}",
      journal = {\apj},
         year = 2019,
        month = feb,
       volume = {872},
       number = {2},
          eid = {144},
        pages = {144},
          doi = {10.3847/1538-4357/aaf7a3},
archivePrefix = {arXiv},
       eprint = {1812.04614},
 primaryClass = {astro-ph.GA},
       adsurl = {https://ui.adsabs.harvard.edu/abs/2019ApJ...872..144H}
}

@ARTICLE{Bundy_2015ApJ...798....7B,
       author = {{Bundy}, Kevin and {Bershady}, Matthew A. and {Law}, David R. and {Yan}, Renbin and {Drory}, Niv and {MacDonald}, Nicholas and {Wake}, David A. and {Cherinka}, Brian and {S{\'a}nchez-Gallego}, Jos{\'e} R. and {Weijmans}, Anne-Marie and {Thomas}, Daniel and {Tremonti}, Christy and {Masters}, Karen and {Coccato}, Lodovico and {Diamond-Stanic}, Aleksandar M. and {Arag{\'o}n-Salamanca}, Alfonso and {Avila-Reese}, Vladimir and {Badenes}, Carles and {Falc{\'o}n-Barroso}, J{\'e}sus and {Belfiore}, Francesco and {Bizyaev}, Dmitry and {Blanc}, Guillermo A. and {Bland-Hawthorn}, Joss and {Blanton}, Michael R. and {Brownstein}, Joel R. and {Byler}, Nell and {Cappellari}, Michele and {Conroy}, Charlie and {Dutton}, Aaron A. and {Emsellem}, Eric and {Etherington}, James and {Frinchaboy}, Peter M. and {Fu}, Hai and {Gunn}, James E. and {Harding}, Paul and {Johnston}, Evelyn J. and {Kauffmann}, Guinevere and {Kinemuchi}, Karen and {Klaene}, Mark A. and {Knapen}, Johan H. and {Leauthaud}, Alexie and {Li}, Cheng and {Lin}, Lihwai and {Maiolino}, Roberto and {Malanushenko}, Viktor and {Malanushenko}, Elena and {Mao}, Shude and {Maraston}, Claudia and {McDermid}, Richard M. and {Merrifield}, Michael R. and {Nichol}, Robert C. and {Oravetz}, Daniel and {Pan}, Kaike and {Parejko}, John K. and {Sanchez}, Sebastian F. and {Schlegel}, David and {Simmons}, Audrey and {Steele}, Oliver and {Steinmetz}, Matthias and {Thanjavur}, Karun and {Thompson}, Benjamin A. and {Tinker}, Jeremy L. and {van den Bosch}, Remco C.~E. and {Westfall}, Kyle B. and {Wilkinson}, David and {Wright}, Shelley and {Xiao}, Ting and {Zhang}, Kai},
        title = "{Overview of the SDSS-IV MaNGA Survey: Mapping nearby Galaxies at Apache Point Observatory}",
      journal = {\apj},
         year = 2015,
        month = jan,
       volume = {798},
       number = {1},
          eid = {7},
        pages = {7},
          doi = {10.1088/0004-637X/798/1/7},
archivePrefix = {arXiv},
       eprint = {1412.1482},
 primaryClass = {astro-ph.GA},
       adsurl = {https://ui.adsabs.harvard.edu/abs/2015ApJ...798....7B}
}

@ARTICLE{Wang_2022ApJ...938L..16W,
       author = {{Wang}, Xin and {Jones}, Tucker and {Vulcani}, Benedetta and {Treu}, Tommaso and {Morishita}, Takahiro and {Roberts-Borsani}, Guido and {Malkan}, Matthew A. and {Henry}, Alaina and {Brammer}, Gabriel and {Strait}, Victoria and {Brada{\v{c}}}, Maru{\v{s}}a and {Boyett}, Kristan and {Calabr{\`o}}, Antonello and {Castellano}, Marco and {Fontana}, Adriano and {Glazebrook}, Karl and {Kelly}, Patrick L. and {Leethochawalit}, Nicha and {Marchesini}, Danilo and {Santini}, P. and {Trenti}, M. and {Yang}, Lilan},
        title = "{Early Results from GLASS-JWST. IV. Spatially Resolved Metallicity in a Low-mass z   3 Galaxy with NIRISS}",
      journal = {\apjl},
         year = 2022,
        month = oct,
       volume = {938},
       number = {2},
          eid = {L16},
        pages = {L16},
          doi = {10.3847/2041-8213/ac959e},
archivePrefix = {arXiv},
       eprint = {2207.13113},
 primaryClass = {astro-ph.GA},
       adsurl = {https://ui.adsabs.harvard.edu/abs/2022ApJ...938L..16W}
}

@ARTICLE{Howk_2018ApJ...856..166H,
       author = {{Howk}, J. Christopher and {Rueff}, Katherine M. and {Lehner}, Nicolas and {Wotta}, Christopher B. and {Croxall}, Kevin and {Savage}, Blair D.},
        title = "{Extraplanar H II Regions in Spiral Galaxies. I. Low-metallicity Gas Accreting through the Disk-halo Interface of NGC 4013}",
      journal = {\apj},
         year = 2018,
        month = apr,
       volume = {856},
       number = {2},
          eid = {166},
        pages = {166},
          doi = {10.3847/1538-4357/aab1fa},
archivePrefix = {arXiv},
       eprint = {1802.10165},
 primaryClass = {astro-ph.GA},
       adsurl = {https://ui.adsabs.harvard.edu/abs/2018ApJ...856..166H}
}

@ARTICLE{Ju_2022ApJ...938...96J,
       author = {{Ju}, Mengting and {Yin}, Jun and {Liu}, Rongrong and {Hao}, Lei and {Shao}, Zhengyi and {Feng}, Shuai and {Riffel}, Rog{\'e}rio and {Liu}, Chenxu and {Stark}, David V. and {Shen}, Shiyin and {Telles}, Eduardo and {Fern{\'a}ndez-Trincado}, Jos{\'e} G. and {Wang}, Junfeng and {Xu}, Haiguang and {Bizyaev}, Dmitry and {Rong}, Yu},
        title = "{MaNGA 8313-1901: Gas Accretion Observed in a Blue Compact Dwarf Galaxy?}",
      journal = {\apj},
         year = 2022,
        month = oct,
       volume = {938},
       number = {2},
          eid = {96},
        pages = {96},
          doi = {10.3847/1538-4357/ac9056},
archivePrefix = {arXiv},
       eprint = {2209.03298},
 primaryClass = {astro-ph.GA},
       adsurl = {https://ui.adsabs.harvard.edu/abs/2022ApJ...938...96J}
}

@ARTICLE{Luo_2021ApJ...908..183L,
       author = {{Luo}, Yuanze and {Heckman}, Timothy and {Hwang}, Hsiang-Chih and {Rowlands}, Kate and {S{\'a}nchez-Menguiano}, Laura and {Riffel}, Rog{\'e}rio and {Bizyaev}, Dmitry and {Andrews}, Brett H. and {Fern{\'a}ndez-Trincado}, Jos{\'e} G. and {Drory}, Niv and {S{\'a}nchez Almeida}, Jorge and {Maiolino}, Roberto and {Lane}, Richard R. and {Argudo-Fern{\'a}ndez}, Maria},
        title = "{Evidence for the Accretion of Gas in Star-forming Galaxies: High N/O Abundances in Regions of Anomalously Low Metallicity}",
      journal = {\apj},
         year = 2021,
        month = feb,
       volume = {908},
       number = {2},
          eid = {183},
        pages = {183},
          doi = {10.3847/1538-4357/abd1df},
archivePrefix = {arXiv},
       eprint = {2012.04073},
 primaryClass = {astro-ph.GA},
       adsurl = {https://ui.adsabs.harvard.edu/abs/2021ApJ...908..183L}
}

@ARTICLE{Long_2025ApJ...986...31L,
       author = {{Long}, Liuze and {Gao}, Yulong and {Gu}, Qiusheng and {Shi}, Yong and {Li}, Xin and {Xu}, Can and {Jin}, Yifei and {Zheng}, Zhiyuan and {Dou}, Jing and {Bian}, Fuyan and {Yu}, Xiaoling},
        title = "{Clumpy Starburst in a Local Dwarf Galaxy, NGC 1522}",
      journal = {\apj},
         year = 2025,
        month = jun,
       volume = {986},
       number = {1},
          eid = {31},
        pages = {31},
          doi = {10.3847/1538-4357/add0b7},
archivePrefix = {arXiv},
       eprint = {2505.10078},
 primaryClass = {astro-ph.GA},
       adsurl = {https://ui.adsabs.harvard.edu/abs/2025ApJ...986...31L}
}

@ARTICLE{Grossi_2025arXiv250118498G,
       author = {{Grossi}, M. and {Gon{\c{c}}alves}, D.~R. and {Krabbe}, A.~C. and {Guti{\'e}rrez Soto}, L.~A. and {Telles}, E. and {Ribeiro}, L.~S. and {Signorini Gon{\c{c}}alves}, T. and {Lopes}, A.~R. and {Smith Castelli}, A.~V. and {De Rossi}, M.~E. and {Lima-Dias}, C. and {Limberg}, G. and {Ferreira Lopes}, C.~E. and {Hernandez-Jimenez}, J.~A. and {Humire}, P.~K. and {Chies Santos}, A.~L. and {Lomel{\'\i}-Nu{\~n}ez}, L. and {Torres-Flores}, S. and {Herpich}, F.~R. and {Oliveira Schwarz}, G.~B. and {Kanaan}, A. and {Mendes de Oliveira}, C. and {Ribeiro}, T. and {Schoenell}, W.},
        title = "{The Southern Photometrical Local Universe Survey (S-PLUS): searching for metal-poor dwarf galaxies}",
      journal = {arXiv e-prints},
         year = 2025,
        month = jan,
          eid = {arXiv:2501.18498},
        pages = {arXiv:2501.18498},
          doi = {10.48550/arXiv.2501.18498},
archivePrefix = {arXiv},
       eprint = {2501.18498},
 primaryClass = {astro-ph.GA},
       adsurl = {https://ui.adsabs.harvard.edu/abs/2025arXiv250118498G}
}

@ARTICLE{Borthakur_2024arXiv240912554B,
       author = {{Borthakur}, Sanchayeeta and {Padave}, Mansi and {Heckman}, Timothy and {Gim}, Hansung B. and {Olvera}, Alejandro J. and {Koplitz}, Brad and {Momjian}, Emmanuel and {Jansen}, Rolf A. and {Thilker}, David and {Kauffman}, Guinevere and {Fox}, Andrew J. and {Tumlinson}, Jason and {Kennicutt}, Robert C. and {Nelson}, Dylan and {Monckiewicz}, Jacqueline and {Naab}, Thorsten},
        title = "{DIISC Survey: Deciphering the Interplay Between the Interstellar Medium, Stars, and the Circumgalactic Medium Survey}",
      journal = {arXiv e-prints},
         year = 2024,
        month = sep,
          eid = {arXiv:2409.12554},
        pages = {arXiv:2409.12554},
          doi = {10.48550/arXiv.2409.12554},
archivePrefix = {arXiv},
       eprint = {2409.12554},
 primaryClass = {astro-ph.GA},
       adsurl = {https://ui.adsabs.harvard.edu/abs/2024arXiv240912554B}
}

@ARTICLE{DiTeodoro_2015MNRAS.451.3021D,
       author = {{Di Teodoro}, E.~M. and {Fraternali}, F.},
        title = "{$^{3D}$ BAROLO: a new 3D algorithm to derive rotation curves of galaxies}",
      journal = {\mnras},
         year = 2015,
        month = aug,
       volume = {451},
       number = {3},
        pages = {3021-3033},
          doi = {10.1093/mnras/stv1213},
archivePrefix = {arXiv},
       eprint = {1505.07834},
 primaryClass = {astro-ph.GA},
       adsurl = {https://ui.adsabs.harvard.edu/abs/2015MNRAS.451.3021D}
}

@ARTICLE{Tinsley_1973ApJ...186...35T,
       author = {{Tinsley}, Beatrice M.},
        title = "{Analytical Approximations to the Evolution of Galaxies}",
      journal = {\apj},
         year = 1973,
        month = nov,
       volume = {186},
        pages = {35-49},
          doi = {10.1086/152476},
       adsurl = {https://ui.adsabs.harvard.edu/abs/1973ApJ...186...35T}
}

@ARTICLE{Maeder_1992A&A...264..105M,
       author = {{Maeder}, Andre},
        title = "{Stellar yields as a function of initial metallicity and mass limit for black hole formation}",
      journal = {\aap},
         year = 1992,
        month = oct,
       volume = {264},
       number = {1},
        pages = {105-120},
       adsurl = {https://ui.adsabs.harvard.edu/abs/1992A&A...264..105M}
}

@ARTICLE{Bouche_2010ApJ...718.1001B,
       author = {{Bouch{\'e}}, N. and {Dekel}, A. and {Genzel}, R. and {Genel}, S. and {Cresci}, G. and {F{\"o}rster Schreiber}, N.~M. and {Shapiro}, K.~L. and {Davies}, R.~I. and {Tacconi}, L.},
        title = "{The Impact of Cold Gas Accretion Above a Mass Floor on Galaxy Scaling Relations}",
      journal = {\apj},
         year = 2010,
        month = aug,
       volume = {718},
       number = {2},
        pages = {1001-1018},
          doi = {10.1088/0004-637X/718/2/1001},
archivePrefix = {arXiv},
       eprint = {0912.1858},
 primaryClass = {astro-ph.CO},
       adsurl = {https://ui.adsabs.harvard.edu/abs/2010ApJ...718.1001B}
}

@ARTICLE{Lilly_2013ApJ...772..119L,
       author = {{Lilly}, Simon J. and {Carollo}, C. Marcella and {Pipino}, Antonio and {Renzini}, Alvio and {Peng}, Yingjie},
        title = "{Gas Regulation of Galaxies: The Evolution of the Cosmic Specific Star Formation Rate, the Metallicity-Mass-Star-formation Rate Relation, and the Stellar Content of Halos}",
      journal = {\apj},
         year = 2013,
        month = aug,
       volume = {772},
       number = {2},
          eid = {119},
        pages = {119},
          doi = {10.1088/0004-637X/772/2/119},
archivePrefix = {arXiv},
       eprint = {1303.5059},
 primaryClass = {astro-ph.CO},
       adsurl = {https://ui.adsabs.harvard.edu/abs/2013ApJ...772..119L}
}

@ARTICLE{Semenov_2021ApJ...918...13S,
       author = {{Semenov}, Vadim A. and {Kravtsov}, Andrey V. and {Gnedin}, Nickolay Y.},
        title = "{Spatial Decorrelation of Young Stars and Dense Gas as a Probe of the Star Formation-Feedback Cycle in Galaxies}",
      journal = {\apj},
         year = 2021,
        month = sep,
       volume = {918},
       number = {1},
          eid = {13},
        pages = {13},
          doi = {10.3847/1538-4357/ac0a77},
archivePrefix = {arXiv},
       eprint = {2103.13406},
 primaryClass = {astro-ph.GA},
       adsurl = {https://ui.adsabs.harvard.edu/abs/2021ApJ...918...13S}
}

@ARTICLE{Schinnerer_2019ApJ...887...49S,
       author = {{Schinnerer}, Eva and {Hughes}, Annie and {Leroy}, Adam and {Groves}, Brent and {Blanc}, Guillermo A. and {Kreckel}, Kathryn and {Bigiel}, Frank and {Chevance}, M{\'e}lanie and {Dale}, Daniel and {Emsellem}, Eric and {Faesi}, Christopher and {Glover}, Simon and {Grasha}, Kathryn and {Henshaw}, Jonathan and {Hygate}, Alexander and {Kruijssen}, J.~M. Diederik and {Meidt}, Sharon and {Pety}, Jerome and {Querejeta}, Miguel and {Rosolowsky}, Erik and {Saito}, Toshiki and {Schruba}, Andreas and {Sun}, Jiayi and {Utomo}, Dyas},
        title = "{The Gas-Star Formation Cycle in Nearby Star-forming Galaxies. I. Assessment of Multi-scale Variations}",
      journal = {\apj},
         year = 2019,
        month = dec,
       volume = {887},
       number = {1},
          eid = {49},
        pages = {49},
          doi = {10.3847/1538-4357/ab50c2},
archivePrefix = {arXiv},
       eprint = {1910.10520},
 primaryClass = {astro-ph.GA},
       adsurl = {https://ui.adsabs.harvard.edu/abs/2019ApJ...887...49S}
}

@software{Bradley_larry_bradley_2023_7946442,
  author       = {Larry Bradley},
  title        = {astropy/photutils: 1.8.0},
  month        = may,
  year         = 2023,
  publisher    = {Zenodo},
  version      = {1.8.0},
  doi          = {10.5281/zenodo.7946442},
  url          = {https://doi.org/10.5281/zenodo.7946442},
}

@ARTICLE{Helfer_2003ApJS..145..259H,
       author = {{Helfer}, Tamara T. and {Thornley}, Michele D. and {Regan}, Michael W. and {Wong}, Tony and {Sheth}, Kartik and {Vogel}, Stuart N. and {Blitz}, Leo and {Bock}, Douglas C. -J.},
        title = "{The BIMA Survey of Nearby Galaxies (BIMA SONG). II. The CO Data}",
      journal = {\apjs},
         year = 2003,
        month = apr,
       volume = {145},
       number = {2},
        pages = {259-327},
          doi = {10.1086/346076},
archivePrefix = {arXiv},
       eprint = {astro-ph/0304294},
 primaryClass = {astro-ph},
       adsurl = {https://ui.adsabs.harvard.edu/abs/2003ApJS..145..259H}
}

@ARTICLE{Vincenzo_Yields_2016MNRAS.455.4183V,
       author = {{Vincenzo}, F. and {Matteucci}, F. and {Belfiore}, F. and {Maiolino}, R.},
        title = "{Modern yields per stellar generation: the effect of the IMF}",
      journal = {\mnras},
         year = 2016,
        month = feb,
       volume = {455},
       number = {4},
        pages = {4183-4190},
          doi = {10.1093/mnras/stv2598},
archivePrefix = {arXiv},
       eprint = {1503.08300},
 primaryClass = {astro-ph.GA},
       adsurl = {https://ui.adsabs.harvard.edu/abs/2016MNRAS.455.4183V}
}

@ARTICLE{Barrera-Ballesteros_2018ApJ...852...74B,
       author = {{Barrera-Ballesteros}, J.~K. and {Heckman}, T. and {S{\'a}nchez}, S.~F. and {Zakamska}, N.~L. and {Cleary}, J. and {Zhu}, G. and {Brinkmann}, J. and {Drory}, N. and {THE MaNGA TEAM}},
        title = "{SDSS-IV MaNGA: What Shapes the Distribution of Metals in Galaxies? Exploring the Roles of the Local Gas Fraction and Escape Velocity}",
      journal = {\apj},
         year = 2018,
        month = jan,
       volume = {852},
       number = {2},
          eid = {74},
        pages = {74},
          doi = {10.3847/1538-4357/aa9b31},
archivePrefix = {arXiv},
       eprint = {1711.10492},
 primaryClass = {astro-ph.GA},
       adsurl = {https://ui.adsabs.harvard.edu/abs/2018ApJ...852...74B}
}

@ARTICLE{Edmunds_1990MNRAS.246..678E,
       author = {{Edmunds}, M.~G.},
        title = "{General Constraints on the Effect of Gas Flows in the Chemical Evolution of Galaxies}",
      journal = {\mnras},
         year = 1990,
        month = oct,
       volume = {246},
        pages = {678},
       adsurl = {https://ui.adsabs.harvard.edu/abs/1990MNRAS.246..678E}
}

@ARTICLE{Broeils_1997A&A...324..877B,
       author = {{Broeils}, A.~H. and {Rhee}, M.-H.},
        title = "{Short 21-cm WSRT observations of spiral and irregular galaxies. HI properties.}",
      journal = {\aap},
         year = 1997,
        month = aug,
       volume = {324},
        pages = {877-887},
       adsurl = {https://ui.adsabs.harvard.edu/abs/1997A&A...324..877B}
}

@ARTICLE{Swaters_2002A&A...390..829S,
       author = {{Swaters}, R.~A. and {van Albada}, T.~S. and {van der Hulst}, J.~M. and {Sancisi}, R.},
        title = "{The Westerbork HI survey of spiral and irregular galaxies. I. HI imaging of late-type dwarf galaxies}",
      journal = {\aap},
         year = 2002,
        month = aug,
       volume = {390},
        pages = {829-861},
          doi = {10.1051/0004-6361:20011755},
archivePrefix = {arXiv},
       eprint = {astro-ph/0204525},
 primaryClass = {astro-ph},
       adsurl = {https://ui.adsabs.harvard.edu/abs/2002A&A...390..829S}
}

@ARTICLE{Noordermeer_2005A&A...442..137N,
       author = {{Noordermeer}, E. and {van der Hulst}, J.~M. and {Sancisi}, R. and {Swaters}, R.~A. and {van Albada}, T.~S.},
        title = "{The Westerbork HI survey of spiral and irregular galaxies. III. HI observations of early-type disk galaxies}",
      journal = {\aap},
         year = 2005,
        month = oct,
       volume = {442},
       number = {1},
        pages = {137-157},
          doi = {10.1051/0004-6361:20053172},
archivePrefix = {arXiv},
       eprint = {astro-ph/0508319},
 primaryClass = {astro-ph},
       adsurl = {https://ui.adsabs.harvard.edu/abs/2005A&A...442..137N}
}

@ARTICLE{Dekel_2009Natur.457..451D,
       author = {{Dekel}, A. and {Birnboim}, Y. and {Engel}, G. and {Freundlich}, J. and {Goerdt}, T. and {Mumcuoglu}, M. and {Neistein}, E. and {Pichon}, C. and {Teyssier}, R. and {Zinger}, E.},
        title = "{Cold streams in early massive hot haloes as the main mode of galaxy formation}",
      journal = {\nat},
         year = 2009,
        month = jan,
       volume = {457},
       number = {7228},
        pages = {451-454},
          doi = {10.1038/nature07648},
archivePrefix = {arXiv},
       eprint = {0808.0553},
 primaryClass = {astro-ph},
       adsurl = {https://ui.adsabs.harvard.edu/abs/2009Natur.457..451D}
}

@ARTICLE{Keres_2009MNRAS.395..160K,
       author = {{Kere{\v{s}}}, Du{\v{s}}an and {Katz}, Neal and {Fardal}, Mark and {Dav{\'e}}, Romeel and {Weinberg}, David H.},
        title = "{Galaxies in a simulated {\ensuremath{\Lambda}}CDM Universe - I. Cold mode and hot cores}",
      journal = {\mnras},
         year = 2009,
        month = may,
       volume = {395},
       number = {1},
        pages = {160-179},
          doi = {10.1111/j.1365-2966.2009.14541.x},
archivePrefix = {arXiv},
       eprint = {0809.1430},
 primaryClass = {astro-ph},
       adsurl = {https://ui.adsabs.harvard.edu/abs/2009MNRAS.395..160K}
}

@ARTICLE{Stewart_2011ApJ...738...39S,
       author = {{Stewart}, Kyle R. and {Kaufmann}, Tobias and {Bullock}, James S. and {Barton}, Elizabeth J. and {Maller}, Ariyeh H. and {Diemand}, J{\"u}rg and {Wadsley}, James},
        title = "{Orbiting Circumgalactic Gas as a Signature of Cosmological Accretion}",
      journal = {\apj},
         year = 2011,
        month = sep,
       volume = {738},
       number = {1},
          eid = {39},
        pages = {39},
          doi = {10.1088/0004-637X/738/1/39},
archivePrefix = {arXiv},
       eprint = {1103.4388},
 primaryClass = {astro-ph.CO},
       adsurl = {https://ui.adsabs.harvard.edu/abs/2011ApJ...738...39S}
}

@ARTICLE{Sankar_2025arXiv251103793S,
       author = {{Sankar}, Sriram and {Stern}, Jonathan and {Power}, Chris and {Catinella}, Barbara and {Fielding}, Drummond and {Faucher-Gigu{\`e}re}, Claude-Andr{\'e} and {Sultan}, Imran and {Boylan-Kolchin}, Michael and {Bland-Hawthorn}, Joss},
        title = "{Hot accretion onto spiral galaxies: the origin of extended and warped HI discs}",
      journal = {arXiv e-prints},
         year = 2025,
        month = nov,
          eid = {arXiv:2511.03793},
        pages = {arXiv:2511.03793},
          doi = {10.48550/arXiv.2511.03793},
archivePrefix = {arXiv},
       eprint = {2511.03793},
 primaryClass = {astro-ph.GA},
       adsurl = {https://ui.adsabs.harvard.edu/abs/2025arXiv251103793S}
}

@ARTICLE{Fraternali_2012MNRAS.426.2166F,
       author = {{Fraternali}, Filippo and {Tomassetti}, Matteo},
        title = "{Estimating gas accretion in disc galaxies using the Kennicutt-Schmidt law}",
      journal = {\mnras},
         year = 2012,
        month = nov,
       volume = {426},
       number = {3},
        pages = {2166-2177},
          doi = {10.1111/j.1365-2966.2012.21650.x},
archivePrefix = {arXiv},
       eprint = {1207.0093},
 primaryClass = {astro-ph.CO},
       adsurl = {https://ui.adsabs.harvard.edu/abs/2012MNRAS.426.2166F}
}

@ARTICLE{Saha_2014MNRAS.444..352S,
       author = {{Saha}, Kanak and {Jog}, Chanda J.},
        title = "{Angular momentum transport and evolution of lopsided galaxies}",
      journal = {\mnras},
         year = 2014,
        month = oct,
       volume = {444},
       number = {1},
        pages = {352-363},
          doi = {10.1093/mnras/stu1414},
archivePrefix = {arXiv},
       eprint = {1407.3349},
 primaryClass = {astro-ph.GA},
       adsurl = {https://ui.adsabs.harvard.edu/abs/2014MNRAS.444..352S}
}

@ARTICLE{Bland-Hawthorn_2017ApJ...849...51B,
       author = {{Bland-Hawthorn}, Joss and {Maloney}, Philip R. and {Stephens}, Alex and {Zovaro}, Anna and {Popping}, Attila},
        title = "{In Search of Cool Flow Accretion onto Galaxies: Where Does the Disk Gas End?}",
      journal = {\apj},
         year = 2017,
        month = nov,
       volume = {849},
       number = {1},
          eid = {51},
        pages = {51},
          doi = {10.3847/1538-4357/aa8f45},
archivePrefix = {arXiv},
       eprint = {1709.08733},
 primaryClass = {astro-ph.GA},
       adsurl = {https://ui.adsabs.harvard.edu/abs/2017ApJ...849...51B}
}

@ARTICLE{Bigiel_2010AJ....140.1194B,
       author = {{Bigiel}, F. and {Leroy}, A. and {Walter}, F. and {Blitz}, L. and {Brinks}, E. and {de Blok}, W.~J.~G. and {Madore}, B.},
        title = "{Extremely Inefficient Star Formation in the Outer Disks of Nearby Galaxies}",
      journal = {\aj},
         year = 2010,
        month = nov,
       volume = {140},
       number = {5},
        pages = {1194-1213},
          doi = {10.1088/0004-6256/140/5/1194},
archivePrefix = {arXiv},
       eprint = {1007.3498},
 primaryClass = {astro-ph.CO},
       adsurl = {https://ui.adsabs.harvard.edu/abs/2010AJ....140.1194B}
}

@ARTICLE{Borthakur_2015ApJ...813...46B,
       author = {{Borthakur}, Sanchayeeta and {Heckman}, Timothy and {Tumlinson}, Jason and {Bordoloi}, Rongmon and {Thom}, Christopher and {Catinella}, Barbara and {Schiminovich}, David and {Dav{\'e}}, Romeel and {Kauffmann}, Guinevere and {Moran}, Sean M. and {Saintonge}, Amelie},
        title = "{Connection between the Circumgalactic Medium and the Interstellar Medium of Galaxies: Results from the COS-GASS Survey}",
      journal = {\apj},
         year = 2015,
        month = nov,
       volume = {813},
       number = {1},
          eid = {46},
        pages = {46},
          doi = {10.1088/0004-637X/813/1/46},
archivePrefix = {arXiv},
       eprint = {1504.01392},
 primaryClass = {astro-ph.GA},
       adsurl = {https://ui.adsabs.harvard.edu/abs/2015ApJ...813...46B}
}

@article{astropy:2013,
Adsurl = {http://adsabs.harvard.edu/abs/2013A%26A...558A..33A},
Archiveprefix = {arXiv},
Author = {{Astropy Collaboration} and {Robitaille}, T.~P. and {Tollerud}, E.~J. and {Greenfield}, P. and {Droettboom}, M. and {Bray}, E. and {Aldcroft}, T. and {Davis}, M. and {Ginsburg}, A. and {Price-Whelan}, A.~M. and {Kerzendorf}, W.~E. and {Conley}, A. and {Crighton}, N. and {Barbary}, K. and {Muna}, D. and {Ferguson}, H. and {Grollier}, F. and {Parikh}, M.~M. and {Nair}, P.~H. and {Unther}, H.~M. and {Deil}, C. and {Woillez}, J. and {Conseil}, S. and {Kramer}, R. and {Turner}, J.~E.~H. and {Singer}, L. and {Fox}, R. and {Weaver}, B.~A. and {Zabalza}, V. and {Edwards}, Z.~I. and {Azalee Bostroem}, K. and {Burke}, D.~J. and {Casey}, A.~R. and {Crawford}, S.~M. and {Dencheva}, N. and {Ely}, J. and {Jenness}, T. and {Labrie}, K. and {Lim}, P.~L. and {Pierfederici}, F. and {Pontzen}, A. and {Ptak}, A. and {Refsdal}, B. and {Servillat}, M. and {Streicher}, O.},
Doi = {10.1051/0004-6361/201322068},
Eid = {A33},
Eprint = {1307.6212},
Journal = {\aap},
Month = oct,
Pages = {A33},
Primaryclass = {astro-ph.IM},
Title = {{Astropy: A community Python package for astronomy}},
Volume = 558,
Year = 2013}

@ARTICLE{astropy:2018,
       author = {{Astropy Collaboration} and {Price-Whelan}, A.~M. and
         {Sip{\H{o}}cz}, B.~M. and {G{\"u}nther}, H.~M. and {Lim}, P.~L. and
         {Crawford}, S.~M. and {Conseil}, S. and {Shupe}, D.~L. and
         {Craig}, M.~W. and {Dencheva}, N. and {Ginsburg}, A. and {Vand
        erPlas}, J.~T. and {Bradley}, L.~D. and {P{\'e}rez-Su{\'a}rez}, D. and
         {de Val-Borro}, M. and {Aldcroft}, T.~L. and {Cruz}, K.~L. and
         {Robitaille}, T.~P. and {Tollerud}, E.~J. and {Ardelean}, C. and
         {Babej}, T. and {Bach}, Y.~P. and {Bachetti}, M. and {Bakanov}, A.~V. and
         {Bamford}, S.~P. and {Barentsen}, G. and {Barmby}, P. and
         {Baumbach}, A. and {Berry}, K.~L. and {Biscani}, F. and {Boquien}, M. and
         {Bostroem}, K.~A. and {Bouma}, L.~G. and {Brammer}, G.~B. and
         {Bray}, E.~M. and {Breytenbach}, H. and {Buddelmeijer}, H. and
         {Burke}, D.~J. and {Calderone}, G. and {Cano Rodr{\'\i}guez}, J.~L. and
         {Cara}, M. and {Cardoso}, J.~V.~M. and {Cheedella}, S. and {Copin}, Y. and
         {Corrales}, L. and {Crichton}, D. and {D'Avella}, D. and {Deil}, C. and
         {Depagne}, {\'E}. and {Dietrich}, J.~P. and {Donath}, A. and
         {Droettboom}, M. and {Earl}, N. and {Erben}, T. and {Fabbro}, S. and
         {Ferreira}, L.~A. and {Finethy}, T. and {Fox}, R.~T. and
         {Garrison}, L.~H. and {Gibbons}, S.~L.~J. and {Goldstein}, D.~A. and
         {Gommers}, R. and {Greco}, J.~P. and {Greenfield}, P. and
         {Groener}, A.~M. and {Grollier}, F. and {Hagen}, A. and {Hirst}, P. and
         {Homeier}, D. and {Horton}, A.~J. and {Hosseinzadeh}, G. and {Hu}, L. and
         {Hunkeler}, J.~S. and {Ivezi{\'c}}, {\v{Z}}. and {Jain}, A. and
         {Jenness}, T. and {Kanarek}, G. and {Kendrew}, S. and {Kern}, N.~S. and
         {Kerzendorf}, W.~E. and {Khvalko}, A. and {King}, J. and {Kirkby}, D. and
         {Kulkarni}, A.~M. and {Kumar}, A. and {Lee}, A. and {Lenz}, D. and
         {Littlefair}, S.~P. and {Ma}, Z. and {Macleod}, D.~M. and
         {Mastropietro}, M. and {McCully}, C. and {Montagnac}, S. and
         {Morris}, B.~M. and {Mueller}, M. and {Mumford}, S.~J. and {Muna}, D. and
         {Murphy}, N.~A. and {Nelson}, S. and {Nguyen}, G.~H. and
         {Ninan}, J.~P. and {N{\"o}the}, M. and {Ogaz}, S. and {Oh}, S. and
         {Parejko}, J.~K. and {Parley}, N. and {Pascual}, S. and {Patil}, R. and
         {Patil}, A.~A. and {Plunkett}, A.~L. and {Prochaska}, J.~X. and
         {Rastogi}, T. and {Reddy Janga}, V. and {Sabater}, J. and
         {Sakurikar}, P. and {Seifert}, M. and {Sherbert}, L.~E. and
         {Sherwood-Taylor}, H. and {Shih}, A.~Y. and {Sick}, J. and
         {Silbiger}, M.~T. and {Singanamalla}, S. and {Singer}, L.~P. and
         {Sladen}, P.~H. and {Sooley}, K.~A. and {Sornarajah}, S. and
         {Streicher}, O. and {Teuben}, P. and {Thomas}, S.~W. and
         {Tremblay}, G.~R. and {Turner}, J.~E.~H. and {Terr{\'o}n}, V. and
         {van Kerkwijk}, M.~H. and {de la Vega}, A. and {Watkins}, L.~L. and
         {Weaver}, B.~A. and {Whitmore}, J.~B. and {Woillez}, J. and
         {Zabalza}, V. and {Astropy Contributors}},
        title = "{The Astropy Project: Building an Open-science Project and Status of the v2.0 Core Package}",
      journal = {\aj},
         year = 2018,
        month = sep,
       volume = {156},
       number = {3},
          eid = {123},
        pages = {123},
          doi = {10.3847/1538-3881/aabc4f},
archivePrefix = {arXiv},
       eprint = {1801.02634},
 primaryClass = {astro-ph.IM},
       adsurl = {https://ui.adsabs.harvard.edu/abs/2018AJ....156..123A}
}

@ARTICLE{astropy:2022,
       author = {{Astropy Collaboration} and {Price-Whelan}, Adrian M. and {Lim}, Pey Lian and {Earl}, Nicholas and {Starkman}, Nathaniel and {Bradley}, Larry and {Shupe}, David L. and {Patil}, Aarya A. and {Corrales}, Lia and {Brasseur}, C.~E. and {N{"o}the}, Maximilian and {Donath}, Axel and {Tollerud}, Erik and {Morris}, Brett M. and {Ginsburg}, Adam and {Vaher}, Eero and {Weaver}, Benjamin A. and {Tocknell}, James and {Jamieson}, William and {van Kerkwijk}, Marten H. and {Robitaille}, Thomas P. and {Merry}, Bruce and {Bachetti}, Matteo and {G{"u}nther}, H. Moritz and {Aldcroft}, Thomas L. and {Alvarado-Montes}, Jaime A. and {Archibald}, Anne M. and {B{'o}di}, Attila and {Bapat}, Shreyas and {Barentsen}, Geert and {Baz{'a}n}, Juanjo and {Biswas}, Manish and {Boquien}, M{'e}d{'e}ric and {Burke}, D.~J. and {Cara}, Daria and {Cara}, Mihai and {Conroy}, Kyle E. and {Conseil}, Simon and {Craig}, Matthew W. and {Cross}, Robert M. and {Cruz}, Kelle L. and {D'Eugenio}, Francesco and {Dencheva}, Nadia and {Devillepoix}, Hadrien A.~R. and {Dietrich}, J{"o}rg P. and {Eigenbrot}, Arthur Davis and {Erben}, Thomas and {Ferreira}, Leonardo and {Foreman-Mackey}, Daniel and {Fox}, Ryan and {Freij}, Nabil and {Garg}, Suyog and {Geda}, Robel and {Glattly}, Lauren and {Gondhalekar}, Yash and {Gordon}, Karl D. and {Grant}, David and {Greenfield}, Perry and {Groener}, Austen M. and {Guest}, Steve and {Gurovich}, Sebastian and {Handberg}, Rasmus and {Hart}, Akeem and {Hatfield-Dodds}, Zac and {Homeier}, Derek and {Hosseinzadeh}, Griffin and {Jenness}, Tim and {Jones}, Craig K. and {Joseph}, Prajwel and {Kalmbach}, J. Bryce and {Karamehmetoglu}, Emir and {Ka{l}uszy{'n}ski}, Miko{l}aj and {Kelley}, Michael S.~P. and {Kern}, Nicholas and {Kerzendorf}, Wolfgang E. and {Koch}, Eric W. and {Kulumani}, Shankar and {Lee}, Antony and {Ly}, Chun and {Ma}, Zhiyuan and {MacBride}, Conor and {Maljaars}, Jakob M. and {Muna}, Demitri and {Murphy}, N.~A. and {Norman}, Henrik and {O'Steen}, Richard and {Oman}, Kyle A. and {Pacifici}, Camilla and {Pascual}, Sergio and {Pascual-Granado}, J. and {Patil}, Rohit R. and {Perren}, Gabriel I. and {Pickering}, Timothy E. and {Rastogi}, Tanuj and {Roulston}, Benjamin R. and {Ryan}, Daniel F. and {Rykoff}, Eli S. and {Sabater}, Jose and {Sakurikar}, Parikshit and {Salgado}, Jes{'u}s and {Sanghi}, Aniket and {Saunders}, Nicholas and {Savchenko}, Volodymyr and {Schwardt}, Ludwig and {Seifert-Eckert}, Michael and {Shih}, Albert Y. and {Jain}, Anany Shrey and {Shukla}, Gyanendra and {Sick}, Jonathan and {Simpson}, Chris and {Singanamalla}, Sudheesh and {Singer}, Leo P. and {Singhal}, Jaladh and {Sinha}, Manodeep and {Sip{H{o}}cz}, Brigitta M. and {Spitler}, Lee R. and {Stansby}, David and {Streicher}, Ole and {{{S}}umak}, Jani and {Swinbank}, John D. and {Taranu}, Dan S. and {Tewary}, Nikita and {Tremblay}, Grant R. and {Val-Borro}, Miguel de and {Van Kooten}, Samuel J. and {Vasovi{'c}}, Zlatan and {Verma}, Shresth and {de Miranda Cardoso}, Jos{'e} Vin{'i}cius and {Williams}, Peter K.~G. and {Wilson}, Tom J. and {Winkel}, Benjamin and {Wood-Vasey}, W.~M. and {Xue}, Rui and {Yoachim}, Peter and {Zhang}, Chen and {Zonca}, Andrea and {Astropy Project Contributors}},
        title = "{The Astropy Project: Sustaining and Growing a Community-oriented Open-source Project and the Latest Major Release (v5.0) of the Core Package}",
      journal = {\apj},
         year = 2022,
        month = aug,
       volume = {935},
       number = {2},
          eid = {167},
        pages = {167},
          doi = {10.3847/1538-4357/ac7c74},
archivePrefix = {arXiv},
       eprint = {2206.14220},
 primaryClass = {astro-ph.IM},
       adsurl = {https://ui.adsabs.harvard.edu/abs/2022ApJ...935..167A}
}

@ARTICLE{Perez_2007CSE.....9c..21P,
       author = {{Perez}, Fernando and {Granger}, Brian E.},
        title = "{IPython: A System for Interactive Scientific Computing}",
      journal = {Computing in Science and Engineering},
         year = 2007,
        month = jan,
       volume = {9},
       number = {3},
        pages = {21-29},
          doi = {10.1109/MCSE.2007.53},
       adsurl = {https://ui.adsabs.harvard.edu/abs/2007CSE.....9c..21P}
}

@INCOLLECTION{Kluyver_2016ppap.book...87K,
       author = {{Kluyver}, Thomas and {Ragan-Kelley}, Benjain and {P{\'e}rez}, Fernando and {Granger}, Brian and {Bussonnier}, Matthias and {Frederic}, Jonathan and {Kelley}, Kyle and {Hamrick}, Jessica and {Grout}, Jason and {Corlay}, Sylvain and {Ivanov}, Paul and {Avila}, Dami{\'a}n and {Abdalla}, Safia and {Willing}, Carol and {Jupyter Development Team}},
        title = "{Jupyter Notebooks{\textemdash}a publishing format for reproducible computational workflows}",
    booktitle = {IOS Press},
         year = 2016,
        pages = {87-90},
          doi = {10.3233/978-1-61499-649-1-87},
       adsurl = {https://ui.adsabs.harvard.edu/abs/2016ppap.book...87K}
}

@Article{Hunter:2007,
  Author    = {Hunter, J. D.},
  Title     = {Matplotlib: A 2D graphics environment},
  Journal   = {Computing in Science \& Engineering},
  Volume    = {9},
  Number    = {3},
  Pages     = {90--95},
  publisher = {IEEE COMPUTER SOC},
  doi       = {10.1109/MCSE.2007.55},
  year      = 2007
}

@Article{         harris2020array,
 title         = {Array programming with {NumPy}},
 author        = {Charles R. Harris and K. Jarrod Millman and St{\'{e}}fan J.
                 van der Walt and Ralf Gommers and Pauli Virtanen and David
                 Cournapeau and Eric Wieser and Julian Taylor and Sebastian
                 Berg and Nathaniel J. Smith and Robert Kern and Matti Picus
                 and Stephan Hoyer and Marten H. van Kerkwijk and Matthew
                 Brett and Allan Haldane and Jaime Fern{\'{a}}ndez del
                 R{\'{i}}o and Mark Wiebe and Pearu Peterson and Pierre
                 G{\'{e}}rard-Marchant and Kevin Sheppard and Tyler Reddy and
                 Warren Weckesser and Hameer Abbasi and Christoph Gohlke and
                 Travis E. Oliphant},
 year          = {2020},
 month         = sep,
 journal       = {Nature},
 volume        = {585},
 number        = {7825},
 pages         = {357--362},
 doi           = {10.1038/s41586-020-2649-2},
 publisher     = {Springer Science and Business Media {LLC}},
 url           = {https://doi.org/10.1038/s41586-020-2649-2}
}

@MISC{specutils,
       author = {{Earl}, Nicholas and {Tollerud}, Erik and {O'Steen}, Ricky and {Brechmos} and {Kerzendorf}, Wolfgang and {Busko}, Ivo and {Shaileshahuja} and {D'Avella}, Dan and {Robitaille}, Thomas and {Ginsburg}, Adam and {Homeier}, Derek and {Sip{\H{o}}cz}, Brigitta and {Averbukh}, Jesse and {Lim}, P.~L. and {Tocknell}, James and {Cherinka}, Brian and {Ogaz}, Sara and {Geda}, Robel and {Davies}, James and {G{\"u}nther}, Hans Moritz and {Conroy}, Kyle and {Barbary}, Kyle and {Foster}, Jonathan and {Droettboom}, Michael and {Torres}, Simon and {Bray}, E.~M. and {Casey}, Andy and {Teuben}, Peter and {Crawford}, Steve and {Ferguson}, Henry},
        title = "{astropy/specutils: V1.8.0}",
 howpublished = {Zenodo},
         year = 2022,
        month = aug,
          eid = {10.5281/zenodo.7015214},
          doi = {10.5281/zenodo.7015214},
      version = {v1.8.0},
    publisher = {Zenodo},
       adsurl = {https://ui.adsabs.harvard.edu/abs/2022zndo...7015214E}
}

@ARTICLE{Sabbi_2018ApJS..235...23S,
       author = {{Sabbi}, E. and {Calzetti}, D. and {Ubeda}, L. and {Adamo}, A. and {Cignoni}, M. and {Thilker}, D. and {Aloisi}, A. and {Elmegreen}, B.~G. and {Elmegreen}, D.~M. and {Gouliermis}, D.~A. and et al.},
        title = "{The Resolved Stellar Populations in the LEGUS Galaxies1}",
      journal = {\apjs},
         year = 2018,
        month = mar,
       volume = {235},
       number = {1},
          eid = {23},
        pages = {23},
          doi = {10.3847/1538-4365/aaa8e5},
archivePrefix = {arXiv},
       eprint = {1801.05467},
 primaryClass = {astro-ph.GA},
       adsurl = {https://ui.adsabs.harvard.edu/abs/2018ApJS..235...23S}
}

@ARTICLE{Rix_1995ApJ...447...82R,
       author = {{Rix}, Hans-Walter and {Zaritsky}, Dennis},
        title = "{Nonaxisymmetric Structures in the Stellar Disks of Galaxies}",
      journal = {\apj},
         year = 1995,
        month = jul,
       volume = {447},
        pages = {82},
          doi = {10.1086/175858},
archivePrefix = {arXiv},
       eprint = {astro-ph/9505111},
 primaryClass = {astro-ph},
       adsurl = {https://ui.adsabs.harvard.edu/abs/1995ApJ...447...82R}
}

@ARTICLE{Zaritsky_1997ApJ...477..118Z,
       author = {{Zaritsky}, Dennis and {Rix}, Hans-Walter},
        title = "{Lopsided Spiral Galaxies and a Limit on the Galaxy Accretion Rate}",
      journal = {\apj},
         year = 1997,
        month = mar,
       volume = {477},
       number = {1},
        pages = {118-127},
          doi = {10.1086/303692},
archivePrefix = {arXiv},
       eprint = {astro-ph/9608086},
 primaryClass = {astro-ph},
       adsurl = {https://ui.adsabs.harvard.edu/abs/1997ApJ...477..118Z}
}

@ARTICLE{Baldwin_1980MNRAS.193..313B,
       author = {{Baldwin}, J.~E. and {Lynden-Bell}, D. and {Sancisi}, R.},
        title = "{Lopsided galaxies}",
      journal = {\mnras},
         year = 1980,
        month = oct,
       volume = {193},
        pages = {313-319},
          doi = {10.1093/mnras/193.2.313},
       adsurl = {https://ui.adsabs.harvard.edu/abs/1980MNRAS.193..313B}
}

@ARTICLE{EspinosaPonce_2020MNRAS.494.1622E,
       author = {{Espinosa-Ponce}, C. and {S{\'a}nchez}, S.~F. and {Morisset}, C. and {Barrera-Ballesteros}, J.~K. and {Galbany}, L. and {Garc{\'\i}a-Benito}, R. and {Lacerda}, E.~A.~D. and {Mast}, D.},
        title = "{H II regions in the CALIFA survey: I. catalogue presentation}",
      journal = {\mnras},
         year = 2020,
        month = may,
       volume = {494},
       number = {2},
        pages = {1622-1646},
          doi = {10.1093/mnras/staa782},
archivePrefix = {arXiv},
       eprint = {2003.07865},
 primaryClass = {astro-ph.GA},
       adsurl = {https://ui.adsabs.harvard.edu/abs/2020MNRAS.494.1622E}
}

@ARTICLE{Binney_1992ARA&A..30...51B,
       author = {{Binney}, James},
        title = "{Warps.}",
      journal = {\araa},
         year = 1992,
        month = jan,
       volume = {30},
        pages = {51-74},
          doi = {10.1146/annurev.aa.30.090192.000411},
       adsurl = {https://ui.adsabs.harvard.edu/abs/1992ARA&A..30...51B}
}

@ARTICLE{vandenBergh_1958AJ.....63..492V,
       author = {{van den Bergh}, S.},
        title = "{Cepheids and galactic structure.}",
      journal = {\aj},
         year = 1958,
        month = dec,
       volume = {63},
        pages = {492-496},
          doi = {10.1086/107816},
       adsurl = {https://ui.adsabs.harvard.edu/abs/1958AJ.....63..492V}
}

@ARTICLE{Talbot_1971ApJ...170..409T,
       author = {{Talbot}, Jr., Raymond J. and {Arnett}, W. David},
        title = "{The Evolution of Galaxies. I. Formulation and Mathematical Behavior of the One-Zone Model}",
      journal = {\apj},
         year = 1971,
        month = dec,
       volume = {170},
        pages = {409},
          doi = {10.1086/151228},
       adsurl = {https://ui.adsabs.harvard.edu/abs/1971ApJ...170..409T}
}

@ARTICLE{Jiang_1999MNRAS.303L...7J,
       author = {{Jiang}, Ing-Guey and {Binney}, James},
        title = "{WARPS and cosmic infall}",
      journal = {\mnras},
         year = 1999,
        month = feb,
       volume = {303},
       number = {1},
        pages = {L7-L10},
          doi = {10.1046/j.1365-8711.1999.02333.x},
archivePrefix = {arXiv},
       eprint = {astro-ph/9807161},
 primaryClass = {astro-ph},
       adsurl = {https://ui.adsabs.harvard.edu/abs/1999MNRAS.303L...7J}
}

@ARTICLE{Lopez_2002A&A...386..169L,
       author = {{L{\'o}pez-Corredoira}, M. and {Betancort-Rijo}, J. and {Beckman}, J.~E.},
        title = "{Generation of galactic disc warps due to intergalactic accretion flows onto the disc}",
      journal = {\aap},
         year = 2002,
        month = apr,
       volume = {386},
        pages = {169-186},
          doi = {10.1051/0004-6361:20020229},
archivePrefix = {arXiv},
       eprint = {astro-ph/0202156},
 primaryClass = {astro-ph},
       adsurl = {https://ui.adsabs.harvard.edu/abs/2002A&A...386..169L}
}

@ARTICLE{Ostriker_1989MNRAS.237..785O,
       author = {{Ostriker}, E.~C. and {Binney}, J.~J.},
        title = "{Warped and tilted galactic discs}",
      journal = {\mnras},
         year = 1989,
        month = apr,
       volume = {237},
        pages = {785-798},
          doi = {10.1093/mnras/237.3.785},
       adsurl = {https://ui.adsabs.harvard.edu/abs/1989MNRAS.237..785O}
}

@ARTICLE{Pringle_1992MNRAS.258..811P,
       author = {{Pringle}, J.~E.},
        title = "{A simple approach to the evolution of twisted accretion discs}",
      journal = {\mnras},
         year = 1992,
        month = oct,
       volume = {258},
       number = {4},
        pages = {811-818},
          doi = {10.1093/mnras/258.4.811},
       adsurl = {https://ui.adsabs.harvard.edu/abs/1992MNRAS.258..811P}
}

@ARTICLE{Ogilvie_1999MNRAS.304..557O,
       author = {{Ogilvie}, G.~I.},
        title = "{The non-linear fluid dynamics of a warped accretion disc}",
      journal = {\mnras},
         year = 1999,
        month = apr,
       volume = {304},
       number = {3},
        pages = {557-578},
          doi = {10.1046/j.1365-8711.1999.02340.x},
archivePrefix = {arXiv},
       eprint = {astro-ph/9812073},
 primaryClass = {astro-ph},
       adsurl = {https://ui.adsabs.harvard.edu/abs/1999MNRAS.304..557O}
}

@ARTICLE{Sancisi_2008A&ARv..15..189S,
       author = {{Sancisi}, Renzo and {Fraternali}, Filippo and {Oosterloo}, Tom and {van der Hulst}, Thijs},
        title = "{Cold gas accretion in galaxies}",
      journal = {\aapr},
         year = 2008,
        month = jun,
       volume = {15},
       number = {3},
        pages = {189-223},
          doi = {10.1007/s00159-008-0010-0},
archivePrefix = {arXiv},
       eprint = {0803.0109},
 primaryClass = {astro-ph},
       adsurl = {https://ui.adsabs.harvard.edu/abs/2008A&ARv..15..189S}
}

@ARTICLE{Padave_2024ApJ...960...24P,
       author = {{Padave}, Mansi and {Borthakur}, Sanchayeeta and {Gim}, Hansung B. and {Thilker}, David and {Jansen}, Rolf A. and {Monkiewicz}, Jacqueline and {Kennicutt}, Robert C. and {Kauffmann}, Guinevere and {Fox}, Andrew J. and {Momjian}, Emmanuel and {Heckman}, Timothy},
        title = "{DIISC-III. Signatures of Stellar Disk Growth in Nearby Galaxies}",
      journal = {\apj},
         year = 2024,
        month = jan,
       volume = {960},
       number = {1},
          eid = {24},
        pages = {24},
          doi = {10.3847/1538-4357/ad029b},
archivePrefix = {arXiv},
       eprint = {2310.08482},
 primaryClass = {astro-ph.GA},
       adsurl = {https://ui.adsabs.harvard.edu/abs/2024ApJ...960...24P}
}

@ARTICLE{Vazdekis_2016MNRAS.463.3409V,
       author = {{Vazdekis}, A. and {Koleva}, M. and {Ricciardelli}, E. and {R{\"o}ck}, B. and {Falc{\'o}n-Barroso}, J.},
        title = "{UV-extended E-MILES stellar population models: young components in massive early-type galaxies}",
      journal = {\mnras},
         year = 2016,
        month = dec,
       volume = {463},
       number = {4},
        pages = {3409-3436},
          doi = {10.1093/mnras/stw2231},
archivePrefix = {arXiv},
       eprint = {1612.01187},
 primaryClass = {astro-ph.GA},
       adsurl = {https://ui.adsabs.harvard.edu/abs/2016MNRAS.463.3409V}
}

@ARTICLE{Salpeter_1955ApJ...121..161S,
       author = {{Salpeter}, Edwin E.},
        title = "{The Luminosity Function and Stellar Evolution}",
      journal = {ApJ},
         year = 1955,
       volume = {121},
        pages = {161},
          doi = {10.1086/145971}
}

@ARTICLE{Lisenfeld_2011A&A...534A.102L,
       author = {{Lisenfeld}, U. and {Espada}, D. and {Verdes-Montenegro}, L. and {Kuno}, N. and {Leon}, S. and {Sabater}, J. and {Sato}, N. and {Sulentic}, J. and {Verley}, S. and {Yun}, M.~S.},
        title = "{The AMIGA sample of isolated galaxies. IX. Molecular gas properties}",
      journal = {\aap},
         year = 2011,
        month = oct,
       volume = {534},
          eid = {A102},
        pages = {A102},
          doi = {10.1051/0004-6361/201117056},
archivePrefix = {arXiv},
       eprint = {1108.2130},
 primaryClass = {astro-ph.CO},
       adsurl = {https://ui.adsabs.harvard.edu/abs/2011A&A...534A.102L}
}
\bibliographystyle{aasjournalv7}



\end{document}